%% file: paper.tex
\documentclass{sig-alt-full}
 \pdfoutput=1
\input newPreamble

\usepackage{dblfloatfix}

\title{Leveraging Cloud Data to Mitigate User Experience \\ from `Breaking Bad'}

\author{
\begin{tabular}{c c c}
Nicholas A. James\footnotemark[2] & Arun Kejariwal\footnotemark[3] & David S. Matteson\footnotemark[2]
\end{tabular} \\[1mm]
\begin{tabular}{c c}
\footnotemark[2] Cornell University & \footnotemark[3] Twitter Inc.
\end{tabular}
}

\begin{document}

\maketitle

\input abstract

\input intro

\input background

\input EDM

\input evaluation

\input relWork

\input conclusion

{
\fontsize{5}{0.19cm}%
\selectfont%
\bibliographystyle{unsrt}
\bibliography{compilers,datamining,masterBib}
}

\input appendix

\end{document}

%% file: newPreamble.tex
\usepackage{hyperref}
\usepackage{wasysym}
\usepackage{multicol}
\usepackage{subcaption}
\usepackage{enumerate,pifont,algorithm2e}
\usepackage{amsmath}
\usepackage[table,xcdraw]{xcolor}

\graphicspath{{figs/}}

\newcommand{\utwi}[1]{\mbox{\boldmath $ #1$}}
\newcommand{\Y}{{\utwi{Y}}}
\newcommand{\X}{{\utwi{X}}}
\DeclareMathOperator*{\argmax}{argmax}

\def\figref#1{Figure~\ref{#1}}
\def\secref#1{Section~\ref{#1}}
\def\ssecref#1{subsection~\ref{#1}}
\def\eqnref#1{Equation~\ref{#1}}
\def\algoref#1{Algorithm~\ref{#1}}
\def\tabref#1{Table~\ref{#1}}

\newtheorem{theorem}{Theorem}
\newtheorem{definition}[theorem]{Definition}
\newtheorem{lemma}[theorem]{Lemma}

%% file: abstract.tex
\begin{abstract}

Low latency and high availability of an app or a web service are key, amongst
other factors, to the overall user experience (which in turn directly impacts
the bottomline). Exogenic and/or endogenic factors often give rise to breakouts
in cloud data which makes maintaining high availability and delivering high
performance very challenging. 
Although there exists a large body of prior research in breakout detection,
existing techniques are not suitable for detecting breakouts in cloud data owing
to being not robust in the presence of anomalies. 

To this end, we developed a novel statistical technique to automatically detect
breakouts in cloud data. In particular, the technique employs {\it Energy
Statistics} to detect breakouts in both application as well as system metrics.
Further, the technique uses robust statistical metrics, viz., median, and
estimates the statistical significance of a breakout through a permutation test.
%
To the best of our knowledge, this is the first work which addresses breakout
detection in the presence of anomalies. 

We demonstrate the efficacy of the proposed technique using
\underline{production} data and report Precision, Recall and F-measure measure.
The proposed technique is $3.5\times$ faster than a state-of-the-art
technique for breakout detection and is being currently used on a daily basis at
Twitter. 

\end{abstract}

%

%% file: intro.tex
\vspace*{-1mm}
\section{Introduction} \label{sec:intro}
\vspace*{-1mm}

\noindent 
In a recent report, Mary Meeker from KPCB mentioned that mobile usage continues
to rise reapidly (14\% Y/Y) and mobile usage now accounts for 25\% of the total 
web usage \cite{ITrends2014}. In a similar vein,  Strategy Analytics reported 
that mobile data traffic is expected to to rise by 300\% by 2017 to a peak of 21 
Exabytes, from 5 Exabytes in 2012 \cite{MTraffic}. Growing traffic and user 
engagement directly impacts the performance and availability of an app/website.
To this end, KISSmetrics reported the following \cite{LoadingTime}:

\begin{enumerate}[{\color{blue}\ding{122}}]
\item $73\%$ of mobile internet users say that they have encountered a website that
      was too slow to load.
      \vspace*{-2mm}
\item $38\%$ of mobile internet users say that they have encountered a website that
      was not available.
      \vspace*{-2mm}
\item A 1 second delay in page response can result in a 7\% reduction in
      conversions.
\end{enumerate}

\begin{figure}[!t] 
\centering
\includegraphics[width=0.92\linewidth]{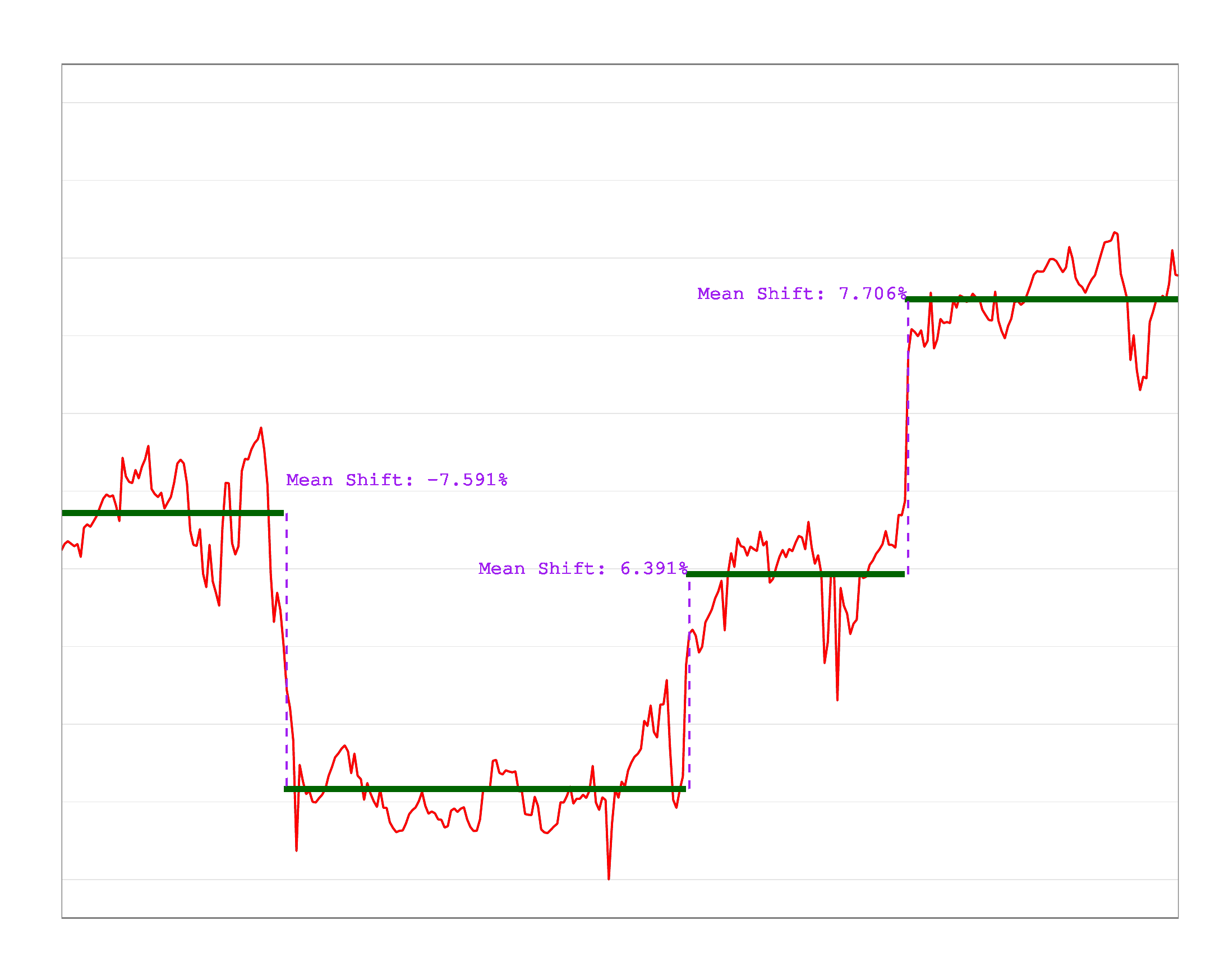}
\vspace*{-4mm}
\caption{Example {\em breakouts} observed in production data at Twitter}
\vspace*{-4mm}
\label{fig:BOs} 
\end{figure}

\noindent 
Likewise, in \cite{LoadingTime1}, it was reported that performance has a direct
impact on business KPIs (Key Performance Indicators). In \cite{LoadingTime2}, 
Shunra (now acquired by Hewlett-Packard) reported: {\sl If your mobile app fails,
48\% of users are less likely to ever use the app again. 34\% of users will just 
switch to a competitor’s app, and 31\% of users will tell friends about their 
poor experience, which eliminates those friends as potential customers}.

Amongst a large multitude of factors, {\em breakouts} -- characterized by either
a mean shift or a rampup from one steady state to another in a given time series
(exemplified in \figref{fig:BOs}) -- in system and/or application metrics can 
potentially impact performance and availability, thereby adversely impacting the
end user experience. A wide variety of factors, some are enumerated below, can 
induce breakouts\footnote{{\em Breakout}, a term commonly used in finance, is 
referred to as {\em changepoint} in statistics.} in system and/or application 
metrics. 

\begin{enumerate}[(a)]
\item Continuous code deployment 
      \vspace*{-1mm}
\item A/B testing \cite{Chatham04,Kohavi09,Siroker13} 
      \vspace*{-1mm}
\item Launch of new products or new product features
      \vspace*{-1mm}
\item Partial failure of a cluster: H\"{o}elzle and Barroso point out that hardware 
      failure in the cloud is more of a norm than exception \cite{Hoelzle09} (also
      see \cite{Dai,Vishwanath10}).
\end{enumerate}

\noindent 
Breakouts can potentially impact latency and availability experienced by the end
user. In light of this, it is critical to detect breakouts early (robust breakout 
detection would also facilitate assessing the efficacy of an A/B test). 
Although there exists a large body of prior research in breakout detection, 
existing techniques are not suitable for detecting breakouts in cloud data owing 
to not being robust in the presence of anomalies. To this end, we developed a 
novel technique to automatically detect breakouts in cloud data (which comprises 
of millions of time series at Twitter \cite{Twitter}). 
The main contributions of the paper are as follows:

\begin{figure}[!t] 
\centering
\includegraphics[width=0.95\linewidth]{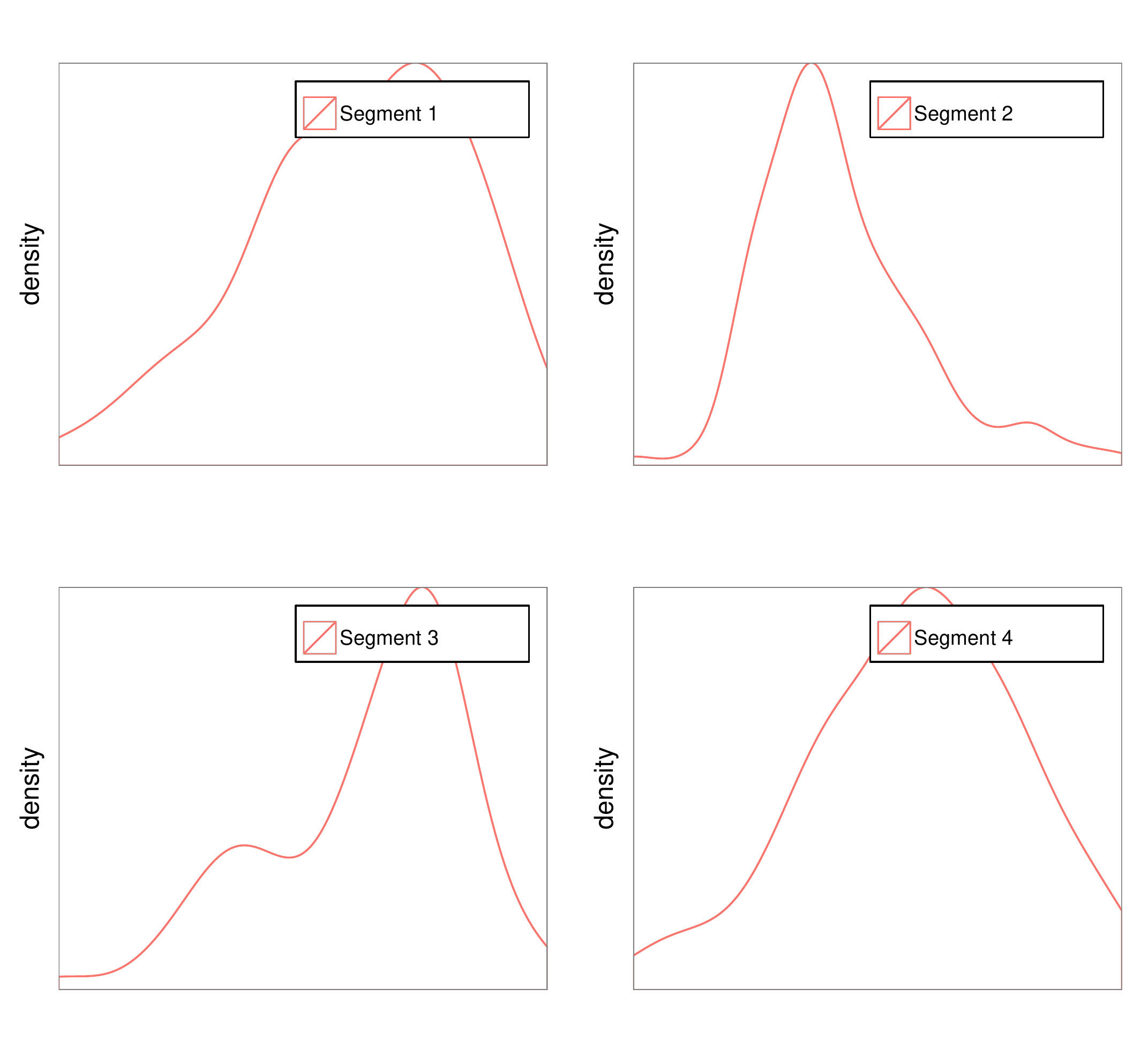}
\vspace*{-4mm}
\caption{Data distribution of the four segments shown in \figref{fig:BOs}}
\vspace*{-4mm}
\label{fig:pdf} 
\end{figure}

\vspace*{-1mm}
\begin{dinglist}{114}
\item First, we propose a novel statistical technique, called {\bf  E-Divisive 
      with Medians (EDM)}, to automatically detect breakouts in cloud data. Unlike 
      the existing techniques for breakout detection, {\bf EDM} is robust against 
      the presence of anomalies.\footnote{Note that the presence of anomalies in 
      production cloud is not uncommon \cite{Vallis:2014}.} 
      The salient features of {\bf EDM} are the following: 
      \begin{dinglist}{122}
      \item {\bf EDM} employs E-statistics \cite{Szekely:2013} to detect divergence 
            in mean. Note that, in general, {\bf EDM} can also be used detect change 
            in distribution in a given time series (discussed further in Section 
            \ref{sec:edm}). 
      \item {\bf EDM} uses robust statistical metrics, viz., median \cite{Huber1981, 
            Hampel1986} , and estimates the statistical significance of a breakout 
            through a permutation test.
      \item {\bf EDM} is non-parametric. This is of paramount importance as the 
            cloud data does not follow the commonly assumed normal distribution, 
            as illustrated by \figref{fig:pdf} or any other widely accepted model. 
            From the figure we note that none of the four segments in \figref{fig:BOs} 
            follow a common distribution.
      \end{dinglist}
      \vspace*{-1mm}
      To the best of our knowledge, this is the first work which addresses breakout 
      detection in the presence of anomalies.
      \vspace*{-1mm}
\item Second, we present a detailed evaluation of {\bf EDM} using production data. 
      \begin{dinglist}{122}
      \item Using production data, we demonstrate that techniques such as {\tt PELT} 
            \cite{Killick:2012} do not fare well when applied to cloud data owing 
            to a non-normal distribution of cloud data. 
      \item We also report Precision, Recall and F-measure to assess the efficacy of
            {\bf EDM}. 
      \end{dinglist}
      \vspace*{-1mm}
      The proposed technique is $3.5\times$ faster than a state-of-the-art
      technique for breakout detection and is being currently used on a daily 
      basis at Twitter.
\end{dinglist}

\noindent
The remainder of the paper is organized as follows: \secref{sec:back} presents
a brief background. \secref{sec:edm} details the proposed technique for
detecting breakouts in cloud data with anomalies. 
\secref{evaluation} presents an evaluation of the proposed technique. 
%
%
Lastly, conclusions and future work are presented in \secref{sec:conclusion}.


%% file: background.tex
\section{Background}\label{sec:back}

\noindent 
In this section we present a brief background of the concepts used by
{\bf EDM} for detecting breakouts. 

\subsection{Divergence Measure}

\noindent 
To detect breakouts, we employ a metric based on the weighted $L^2$-distance
between the characteristic functions of random variables. 
Let $X$ and $Y$ be independent random variables, $X'$ be an i.i.d copy 
of $X$ and $Y'$ be an i.i.d. copy of Y.  Let the cumulative distribution
function of $X$ and $Y$ be denoted by $F$ and $G$ respectively. 

\begin{definition}
The energy distance between $X$ and $Y$ is defined as follows \cite{Szekely:2013}:
\begin{equation}
\mathcal E(X,Y) = 2E|X-Y| - E|X-X'| - E|Y-Y'|
\label{eq:edist}
\end{equation} 
\end{definition}

\noindent 
In \cite{Rizzo:2010}, Rizzo and Sz\'ekely  showed that the $L^2$-distance 
between $F$ and $G$ satisfies the following: 

\begin{equation}
2 \int_{-\infty}^{\infty} (F(x) - G(x))^2dx = \mathcal E(X,Y)
\label{eq:edist1}
\end{equation} 

\noindent 
For a random variable $X$, its characteristic function $\phi_x(t)$ is defined
by $\phi_x(t) = E(\exp\{iXt\})$. Using this notation, Sz\'ekely and Rizzo 
\cite{Szekely:2013} show that the energy distance between $X$ and $Y$ can also 
be represented in terms of their characteristic functions:

\begin{equation}
\mathcal E(X,Y) = \int_{-\infty}^\infty\frac{|\phi_x(t)-\phi_y(t)|^2}{\pi t^2}dt.
\label{distChar}
\end{equation}

\noindent 
Since the characteristic function, like the cumulative distribution function,
uniquely defines a random variable, we define a class of distance measures 
based on them. Let
 
\begin{equation}
\mathcal D(X,Y;\alpha) = \int_{-\infty}^\infty |\phi_x(t)-\phi_y(t)|^2\omega(t;\alpha)dt
\label{L2Dist}
\end{equation}

\noindent 
where $\omega(t;\alpha)$ is a weight function, parameterized by $\alpha$, such
that $D(X,Y;\alpha)<\infty$. The indexing parameter $\alpha$ is used to 
scale the distance between distributions.
For instance, the metric used in Equation \ref{distChar} is obtained by using 
$\omega(t;\alpha) = \frac{1}{\pi t^2}$.
In \cite{Szekely:2005}, Sz\'ekely and Rizzo suggested the following for $\omega$:

\begin{equation*}
\omega(t;\alpha) = \left(\frac{2\pi^{\frac{1}{2}}\Gamma(1-\alpha/2)}{\alpha2^\alpha\Gamma[(1+\alpha)/2]}|t|^\alpha\right)^{-1}
\label{omega}
\end{equation*}

\noindent 
where $\Gamma(\cdot)$ is the complete gamma function. 
Using this weight function allows us to obtain a metric that generalizes the one 
in \eqnref{eq:edist}. For $\alpha \in (0,2]$, the generalized energy distance
between $X$ and $Y$ is given by:
$$
\mathcal E(X,Y;\alpha)=2E|X-Y|^\alpha-E|X-X'|^\alpha-E|Y-Y'|^\alpha
$$
Sz\'ekely and Rizzo \cite{Rizzo:2010} also show that with this weight function, 
and $\alpha\in (0,2]$, we have 
$$
\mathcal D(X,Y;\alpha) = \mathcal E(X,Y;\alpha).
$$

\noindent For detecting divergence in mean, $\alpha$ is set 
to 2; on the other hand, for detecting arbitrary change in distribution, $0 < \alpha < 2$ 
may be a better choice \cite{Szekely:2005}. This property is exemplified through the 
following Lemma.

\setcounter{theorem}{0}
\begin{lemma}
For any pair of independent random variables $X$ and $ Y$, 
and for any $\alpha \in (0,2),$
if $E(|X|^{\alpha} + |Y|^{\alpha}) < \infty,$ 
then ${\mathcal E}(X,Y; \alpha) \in [0,\infty),$ 
and ${\mathcal E}(X,Y; \alpha) = 0$ if and only if $X$ and $Y$ are identically distributed. 
Furthermore, if $\alpha=2$, we have that $\mathcal E(X,Y;2)=0$ if and only if $EX=EY$.
\end{lemma}
\begin{proof}
A proof is given in \cite{Szekely:2005}.
\end{proof}
\setcounter{theorem}{1}

\noindent 
The metric $\mathcal E$ allows for  a simple and intuitive approximation to $\mathcal D$ 
and doesn't require any integration. 
Let $\X_n = \{ X_i : i = 1,\ldots,n \}$ and $\Y_m = \{ Y_j : j = 1,\ldots,m \}$
be independent iid samples from the distribution of $X, Y \in \mathbb{R}^{d},$
respectively, such that $E|X|^{\alpha}, E|Y|^{\alpha} < \infty$ for some $\alpha 
\in (0,2)$. We can then approximate $\mathcal E$ by $\widehat{\mathcal E}$ as 
follows:

\begin{eqnarray}
\widehat{\mathcal E}(\X_n,\Y_m;\alpha) &=& \frac{2}{nm}\sum_{i=1}^n\sum_{j=1}^m|x_i-y_j|^\alpha \nonumber \\
                                       & & -\binom{n}{2}^{-1}\sum_{i<j}|x_i-x_j|^\alpha \\
                                       & & -\binom{m}{2}^{-1}\sum_{i<j}|y_i-y_j|^\alpha \nonumber
\label{eStat}
\end{eqnarray}

\noindent 
The first term on the right hand side of \eqnref{eStat} correspond to the 
{\em between} distance between $\X_n$ and $\Y_m$. The second and third terms on 
the right side of \eqnref{eStat} correspond to the {\em within} distance
of $\X_n$ and $\Y_m$ respectively \cite{Szekely:2005}.   

By the strong law of large numbers for U-statistics \cite{Hoeffding:1961}, 
$\widehat{\mathcal E} \to \mathcal E$ as $\min(n,m)\to\infty$. 
Furthermore, Sz\'ekely and Rizzo \cite{Rizzo:2010} show that under the null hypothesis 
of equal distributions, i.e., $\mathcal E(X,Y;\alpha)=0$, 
$$
\frac{nm}{n+m}\widehat{\mathcal E}(\X_n,\Y_m;\alpha)\Rightarrow\mathcal A
$$
\noindent
as $\min(n,m)\to\infty$, where $\mathcal A$ is a non-degenerate random variable 
and $M \Rightarrow N$ means that $M$ converges in distribution to $N$. 
However, under the alternative hypothesis, $\frac{nm}{n+m}\widehat{\mathcal E}\to\infty$ 
as $\min(m,n)\to\infty$. For notational simplicity, we will use the following 
in the remainder of the paper:
\begin{equation}
\widehat{\mathcal Q}(\X_n,\Y_m;\alpha)=\frac{nm}{n+m}\widehat{\mathcal E}(\X_n,\Y_m;\alpha)
\label{qHat} 
\end{equation}

\subsection{Permutation Test}\label{sec:permTest}

\noindent
The convergence of the statistic presented in Equation \ref{qHat} allows us 
to determine the statistical significance of a proposed breakout. Let the  
observations of a time series be given by 
$Z_1, Z_2,\dots, Z_n$ and $1\le\tau<\kappa\le n$ 
be constants. We define the following sets $A_\tau=\{Z_1, Z_2,\dots, Z_\tau\}$ 
and $B_\tau(\kappa)=\{Z_{\tau+1},\dots, Z_\kappa\}$. A breakout location $\hat\tau$ 
is then estimated as the value that maximizes

\begin{equation*}
\widehat{\mathcal Q}(A_\tau, B_\tau(\kappa); \alpha)
\end{equation*}

\noindent 
for $1\le\tau<\kappa\le n$. Along with the estimated breakout location we also
have an associated test statistic

\begin{equation*}
\hat q = \widehat{\mathcal Q}(A_{\hat\tau},B_{\hat\tau}(\hat\kappa);\alpha).
\end{equation*}

\noindent 
Given $\alpha=2$, large values of $\hat q$ correspond to a significant change in mean (and 
a distribution in general). However, calculating a precise critical value 
requires a knowledge of the underlying distributions, which are generally 
unknown. Therefore, we propose a permutation test to determine the 
significance of $\hat q$.

Under the null hypothesis that there does not exist a breakout, we conduct 
a permutation test as follows. First, the observations are permuted to 
construct a new time series.  Then, we re-apply the estimation procedure to 
the permuted observations. This process is repeated and after the $r$th 
permutation of the observations we record the value of the test statistic 
$\hat q^{(r)}$.

This permutation test will result in an exact p-value if we consider all
possible permutations. However, this is not computationally tractable in 
general. Therefore, we obtain an approximate p-value by performing a 
sequence of $R$ \emph{random} permutations. The approximate p-value is 
computer as follows: 
$$
\#\{r: \hat q^{(r)}\ge\hat q\}/(R+1)
$$ 
\noindent 
The re-sampling risk, the probability of a different decision than the one 
based on the theoretical p-value, can be uniformly bounded by an arbitrarily 
small constant using the approach proposed by Gandy \cite{Gandy09}.
In our analysis we test at the $5\%$ significance level and use $R=199$ 
permutations.


\subsection{Metrics}\label{sec:metrics}

\noindent 
In order to minimize user impact, it is imperative to detect breakout(s)
at the earliest. We qualify the timeliness of breakout detection via {\bf EDM} 
using the metric {\bf TTD} defined below: 

\begin{definition}
We define {\bf TTD (Time to Detect)} as the number of time series 
observations between the occurrence of a breakout and the 
breakout estimate reported by a breakout detection algorithm.
\end{definition}

{\em Precision} is the ratio of true positives (tp) over the sum of true positives 
(tp) and false positives (fp). 
{\em Recall} is the ratio of true positives (tp) over the sum of true positives 
(tp) and false negatives (fn). 
{\em F-measure} is defined as follows (refer to \cite{Pang:2006} for a detailed 
discussion): 

\begin{equation} 
F = 2 \times \frac{\text{precision} \times \text{recall}}{\text{precision} + \text{recall}} 
\label{eq:fm}
\end{equation}


\input{motivateEx.tex}


%% file: motivateEx.tex
\begin{figure*}[!t]
\centering
\hspace*{-6mm}
\includegraphics[width=1.05\textwidth,height=1.8in]{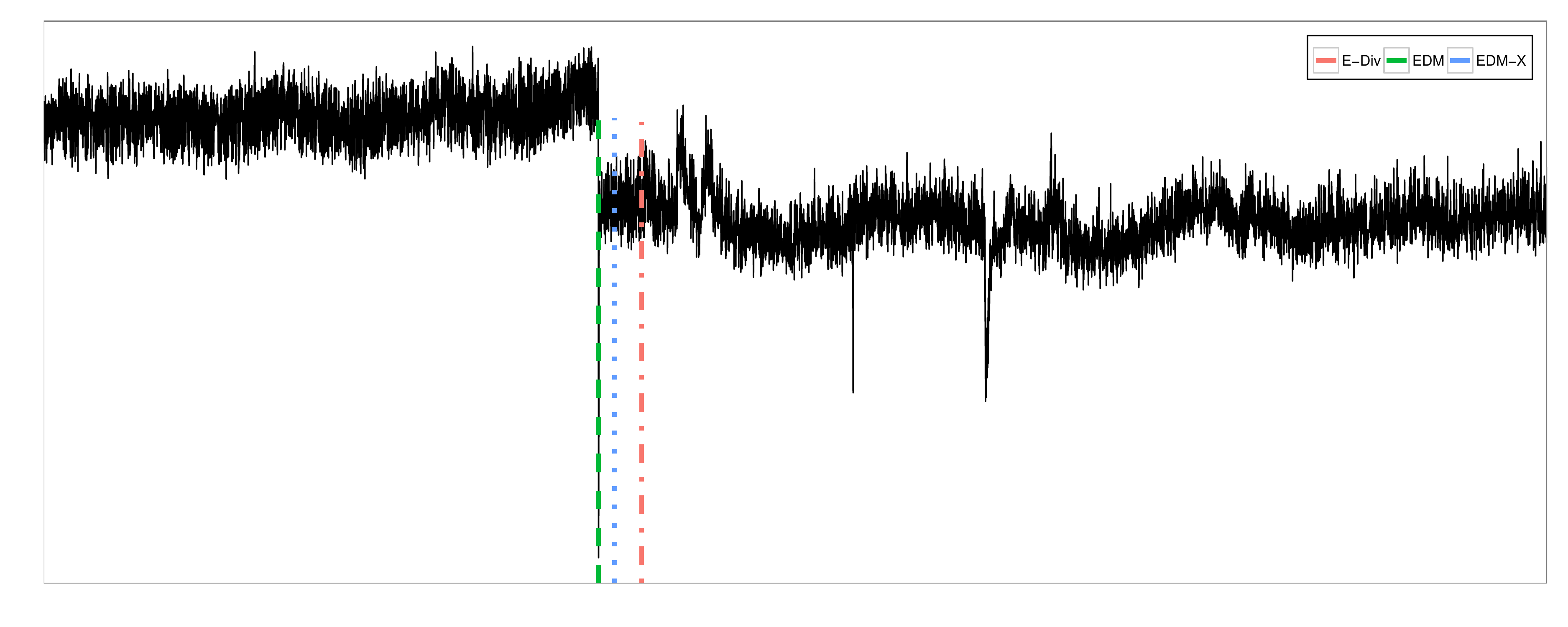}
\vspace*{-8mm}
\caption{Example (using \underline{production} data) highlighting the impact 
         of presence of anomalies on breakout detection}
\vspace*{-6mm}
\label{motiv}
\end{figure*}

%% file: EDM.tex
\section{E-Divisive with Medians} \label{sec:edm}

\noindent 
Suppose that we are given the following time series, $Z_1, Z_2,\dots,$ $Z_n$ 
consisting of independent observations. A breakout is characterized by a 
value $\gamma\in (0,1)$ such that observations $\{Z_1, Z_2,\dots, Z_{\lfloor 
\gamma N\rfloor}\}$ have distribution function $F$, and observations 
$\{Z_{\lfloor \gamma N\rfloor+1}, Z_{\lfloor\gamma N\rfloor +2},\dots, Z_n\}$ 
have distribution function $G$. Furthermore, it is assumed that $F\neq G$. In 
order to determine if the observations in the provided time series are 
identically distributed we perform the following hypothesis test:

\vspace*{-4mm}
\begin{equation*}
\text{H}_0: \gamma=1
\end{equation*}

\vspace*{-4mm}
\begin{equation*}
\text{H}_A: 0<\gamma<1
\end{equation*}
\vspace*{-1mm}

\noindent 
If the null hypothesis of no breakout is rejected, we must then also return an
estimate for the breakout location. Prior work in breakout detection assumes
that the time series under consideration is free of anomalies. However, this is
not the case in \underline{production} cloud data. \figref{motiv} illustrates 
the impact of the presence of anomalies on the location of a breakout detected.
From the figure we note that there are multiple global anomalies, both positive
and negative. The breakout locations obtained using {\tt E-Div} (of the {\tt ecp} 
R package \cite{James:2012}) and the algorithms -- {\bf EDM} and {\bf EDM-X} --
presented later in this section are marked with vertical lines. 
From a TTD perspective, we note that using the proposed algorithms we obtain 
estimates of the location of breakouts than other non-parametric procedures.
This is due to the fact that {\bf EDM} and {\bf EDM-X} are anomaly ``aware". 

\subsection{Robustness against anomalies}

\noindent 
The approximation, $\widehat{\mathcal E}$ given in Equation \ref{eStat} is 
susceptible to anomalies since one single anomaly can greatly change its value. 
This is due to the fact that $\widehat{\mathcal E}$ is based upon a linear 
combination of sample means. To alleviate this issue we instead use a robust 
location estimator, the median. We thus define the robust between sample 
distance:

\vspace*{-4mm}
$$ 
m_{XY}^\alpha = \text{median}\left\{|x_i-y_j|^\alpha: 1\le i\le n, 1\le j\le m\right\}
$$

\noindent
and similarly define $m_{XX}$ and $m_{YY}$ as the median of the within 
sample distances. We then obtain a robust version of $\widehat{\mathcal E}$ 
as follows:

\vspace*{-4mm}
\begin{equation}
\widetilde{\mathcal E}(\X_n,\Y_m;\alpha)=2m_{XY}^\alpha-m_{XX}^\alpha-m_{YY}^\alpha
\label{robustEStat}
\end{equation}

\input{headTail.tex}

\noindent 
For any two given sets $\X_n$ and $\Y_m$ the time necessary to compute 
Equation \ref{robustEStat} a single time is linearly proportional to 
the number of distance terms present. Therefore, if we assume that 
$n\ge m$, both $\widehat{\mathcal E}$ and $\widetilde{\mathcal E}$ 
require $O(n^2)$ calculations to evaluate.\par

However, if a single observation is added/removed from either $\X_n$ or 
$\Y_m$, the value of $\widehat{\mathcal E}$ can be updated in $O(n)$ time, 
but $\widetilde{\cal E}$  will require $O(n^2)$. However, if we use a tree 
data structure, we can update $\widetilde{\cal E}$ in $O(n\log(n))$  time; 
but, this comes at the expense of needing $O(n^2\log(n))$ time to calculate 
the initial value of our statistic. Since such updates may be done a large 
number of times, we consider this trade-off to be acceptable.

Although we can now quickly perform updates we will have to keep track of 
all $O(n^2)$ distances. Even for moderately sized time series this may become 
intractable, even with 24GB of memory. For this reason we make use of interval 
trees (see the Appendix for further details) in order to obtain an approximate 
median. Through experimentation we learned that even the $O(n\log(n))$ update 
is too slow, and thus we use the following approximation.

Let $\delta>1$. We approximate the within distance for the set $\X_n$ as 
follows:

\begin{equation*}
m_{XX}^{\alpha,\delta}=\text{median}\left\{|x_i-x_j|^\alpha: 1\le i<j\le\delta\mbox{ or }i+1=j\right\}
\end{equation*}

\noindent 
We similarly define $m_{YY}^{\alpha,\delta}$. The between distance is 
approximated by using only $\delta$ observations from each set. 
\figref{headTail} shows two possible ways of selecting the $\delta$ 
observations. 

\begin{itemize}
\item[\textbf{Head}] \figref{headTail} (A) chooses to take the $\delta$ 
                     observations that are at head of both sets {\bf X} 
                     and {\bf Y}.
\item[\textbf{Tail}] \figref{headTail} (B) chooses to take the $\delta$ 
                     observations at the tail of set {\bf X} and the head 
                     of {\bf Y}.
\end{itemize}

\noindent 
Based on our experiments using production data we learned that using the
{\bf Tail} (as illustrated in \figref{headTail} (B)) yields better breakout 
estimates and hence, we use: 

\begin{equation*}
m_{XY}^{\alpha,\delta}=\text{median}\left\{|x_i-y_j|^\alpha: n-\delta+1\le i\le n, 1\le j\le \delta\right\}
\end{equation*}

\noindent 
In light of the aforementioned approximation, Equation \ref{qHat} can be written as:

\begin{equation}
\widetilde{\cal Q}(\X_n,\Y_m;\alpha,\delta) = \frac{nm}{n+m}\widetilde{\cal E}(\X_n,\Y_m;\alpha,\delta)
\label{qRobust}
\end{equation}

\noindent 
Typically $\delta$ is chosen such that it is much smaller than $\sqrt{n}$. 
Therefore, with these approximations we can create the statistic 
$\widetilde{\cal E}(\X_n,\Y_n;\alpha,\delta)$, which can be calculated in 
$O(n\log(n))$ time and updated in $O(\log(n))$ time when using the interval 
tree approximation.

\subsection{Algorithm}

\noindent 
The \textbf{EDM} algorithm makes use of the $\widetilde{\cal E}(\cdot,\cdot;\alpha,\delta)$ 
statistic presented in the previous section. Let a time series be given 
by $Z_1, Z_2, \dots, Z_n$ and $1<\delta\le\tau$ and $\tau+\delta\le\kappa\le n$. 
We define the following sets: $A_\tau=\{Z_1, Z_2,\dots, Z_\tau\}$ and 
$B_\tau(\kappa)=\{Z_{\tau+1}, Z_{\tau+2},\dots, Z_\kappa\}$. Thus, both 
$A_\tau$ and $B_\tau(\kappa)$ have at least $\delta$ observations. 
Using Equation \ref{qRobust} we obtain the breakout estimate, $\hat\tau$, 
as follows:

\begin{equation}
(\hat\tau,\hat\kappa) = \argmax_{\tau,\kappa}\ \widetilde{\cal Q}(A_\tau, B_\tau(\kappa); \alpha,\delta)
\label{breakEst}
\end{equation}

\input{edmAlgo.tex}

\noindent 
By solving the maximization problem given in \eqnref{breakEst} we not only 
obtain an estimate $\hat\tau$, but also its associated test statistic 
value $\hat q$. Given this and a predetermined significance
level, we perform a permutation test (detailed in \ssecref{sec:permTest}) 
to determine whether the reported breakout is statistically significant.

Algorithm \ref{algo:edm} is used to determine $\hat\tau$ and $\hat\kappa$. 
We set $D=10$ in our implementation. However, we suggest selecting $D$ such 
that $2^D\approx n$. Then, the algorithm makes use of two key procedures, 
{\tt \ref{algo:forwardUpdate}} and {\tt \ref{algo:backwardUpdate}}. These 
procedures allow us to efficiently update $\widetilde{\mathcal Q}$ by making
use of the current states of the interval trees. 

\input fwd_back_update.tex

\begin{itemize}
\item \ref{algo:forwardUpdate} iterates $\kappa$ from $\tau+\delta+1$ to $n$ 
      and updates the value of $\widetilde{\mathcal Q}$ after each iteration.
      Each iteration corresponds to adding values to $B_\tau(\kappa)$.
\item \ref{algo:backwardUpdate} iterates $\kappa$ from $n-1$ to $\tau+\delta+1$
      and updates the value of $\widetilde{\mathcal Q}$ after each iteration. 
      Each iteration corresponds to removing values from $B_\tau(\kappa)$.
\item For both procedures \ref{algo:forwardUpdate} and \ref{algo:backwardUpdate}, all 
      the parameters are passed by reference. Additionally, both procedures 
      obtain the an approximate medians in $\mathcal O(D)$ (refer to the 
      Appendix for details). 
      In both cases, all the interval trees are updated. Hence, the statistic
      value can be computed in logarithmic time.
\end{itemize}

\input{edmxAlgo.tex}

\subsubsection{\textbf{Special Case: $\alpha=2$}}

\noindent 
It should be noted that when $\alpha=2$, it is possible to obtain a much more 
efficient algorithm. In this case, ${\cal E}(X,Y;2)=2(EX-EY)^2$; hence, changes 
in mean can be detected. As mentioned before, a robust location can be estimate
by considering the sample median instead of the sample mean. In this case, 
we define $\widetilde{\cal E}$ as follows:

\begin{equation*}
\widetilde{\cal E}(A_\tau, B_\tau(\kappa);2,\delta) = 2[\text{median}(A_\tau)-\text{median}(B_\tau(\kappa))]^2
\end{equation*}

\noindent 
This algorithm only considers the median of the actual observations and not
the median of their distances. Unlike the case where $0<\alpha<2$, only $O(n)$ 
additional memory is required and updates can be performed in $O(\log(n))$ time.
This simplification enables the use of exact medians instead of approximations.
Because of this feature, we call this algorithm {\bf E-Divisive with Exact Medians
(EDM-X)} -- see \algoref{algo:edmx}. The algorithm is able to keep track of
the exact medians by using pairs of heaps; a max-heap stores the $\frac{n}{2}$
smallest observations, while a min-heap stores the $\frac{n}{2}$ largest
observations. If $n$ is odd, then one of these heaps will have an additional
element. Procedure {\tt \ref{algo:AddToHeap}} is used by Algorithm
\ref{algo:edmx} to maintain these properties for a given pair of heaps. And since 
the heaps are passed by reference no extra space or time is required to make copies.

%% file: headTail.tex
\begin{figure}[!b]
\vspace*{-3mm}
\centering
\includegraphics[scale=.30]{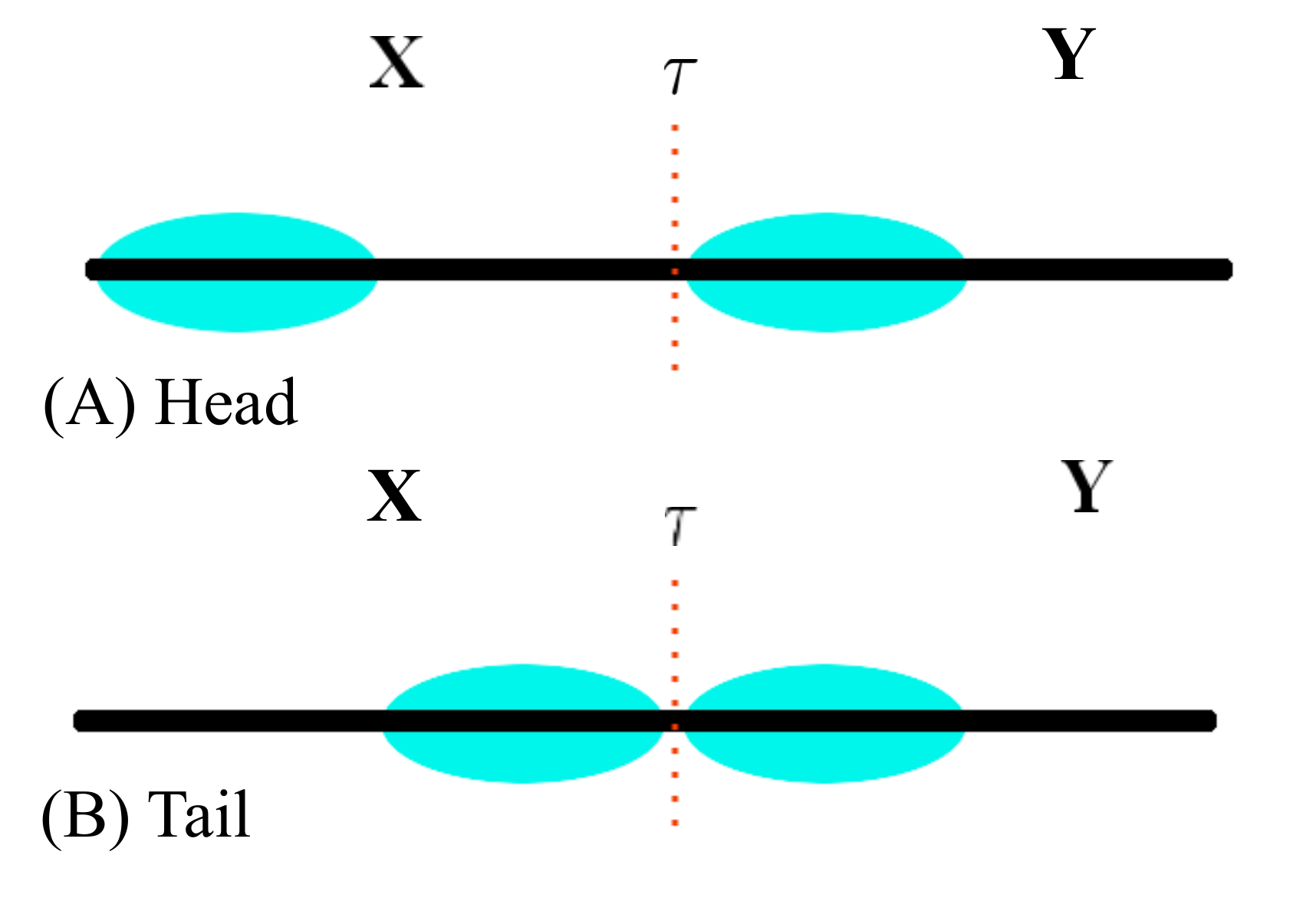}
\vspace*{-5mm}
\caption{
This figure depicts two different ways of selecting which $\delta$ observations to use for approximating the between distance. 
}
\label{headTail}
\end{figure}

%% file: edmAlgo.tex
\begin{algorithm}[!b]
\SetKwInOut{Input}{Parameters}
\SetKwInOut{Output}{Output}
\caption{\textbf{EDM}}
\Input{$Z$, $\delta$, and $D$}
Let $T_A, T_B$, and $T_{AB}$ be interval trees with $2^D$ leaf nodes \\
\tcp{Initialize within distance trees}
\For{$1\le i\le\delta$}{
	\For{$i+1\le j\le\delta$}{
		Insert $|Z_i-Z_j|$ to $T_A$\\
		Insert $|Z_{i+\delta}-Z_{j+\delta}|$ to $T_B$\\
	}
}
\tcp{Initialize between distance tree}
\For{$1\le i\le\delta$}{
	\For{$1\le j\le\delta$}{
		Insert $|Z_i-Z_{j+\delta}|$ to $T_{AB}$\\
	}
}
$\langle m1, m2, m3 \rangle$ = approx. median $\langle T_{AB}$, $T_A$, $T_B \rangle$ \\
$bestStat$ = $\frac{\tau(\kappa-\tau)}{\kappa}(2m1 - m2 - m3)$\\
$bestLoc = \delta$\\
$\tau=\delta$\\
$forwardMove  = 0$\\
\tcp{Update trees} 
\While{$\tau\le n-\delta$}{
	\If{forwardMove = 1}{
		Perform ForwardUpdate\\
	}
	\Else{
		Perform BackwardUpdate\\
	}
	\textit{forwardMove} = 1 - \textit{forwardMove}
}
\Return{$bestLoc$}
\label{algo:edm}
\end{algorithm}

%% file: fwd_back_update.tex
\begin{procedure}[!b]
\SetKwInOut{Input}{Parameters}
\SetKwInOut{Output}{Output}
\caption{ForwardUpdate()}
\Input{$Z$, $\delta$, $T_A, T_B$, $T_{AB}$, $\tau$, \textit{bestStas}, \textit{bestLoc}}
$n = Z.size()$\\
$\tau\leftarrow\tau+1$\\
Update counts in $T_A$, $T_B$, and $T_{AB}$ resulting from new $\tau$ value\\
\For{$\tau+\delta\le\kappa\le n$}{
	Insert |$Z_\kappa-Z_{\kappa-1}|$ to tree $T_B$\\
        $\langle m1, m2, m3 \rangle$ = approx. median $\langle T_{AB}$, $T_A$, $T_B \rangle$ \\
	\textit{stat} = $\frac{\tau(\kappa-\tau)}{\kappa}(2m1 - m2 - m3)$\\
	\If{$stat > bestStat$}{
		$bestStat = stat$\\
		$bestLoc = \tau$\\
	}
}
\label{algo:forwardUpdate}
\end{procedure}

\begin{procedure}[!b]
\SetKwInOut{Input}{Parameters}
\SetKwInOut{Output}{Output}
\caption{BackwardUpdate()}
\Input{$Z$, $\delta$, $T_A, T_B, T_{AB}$, $\tau$, \textit{bestStat}, \textit{bestLoc}}
$n=Z.size()$\\
$\tau\leftarrow\tau+1$\\
Update counts in $T_A, T_B$ and $T_{AB}$ resulting from new $\tau$ value\\
$\kappa = n$\\
\While{$\kappa\ge\tau+\delta$}{
	Insert $|Z_{\kappa}-Z_{\kappa-1}|$ to tree $T_B$\\
        $\langle m1, m2, m3 \rangle$ = approx. median of $\langle T_{AB}$, $T_A$, $T_B \rangle$ \\
	stat = $\frac{\tau(\kappa-\tau)}{\kappa}(2m1 - m2 - m3)$\\
	\If{$stat > bestStat$}{
		$bestStat = stat$\\
		$bestLoc = \tau$\\
	}
	$\kappa\leftarrow\kappa-1$
}
\label{algo:backwardUpdate}
\end{procedure}

%% file: edmxAlgo.tex
\begin{algorithm}[h]
\SetKwInOut{Input}{Parameters}
\SetKwInOut{Output}{Output}
\caption{\textbf{EDM-X}}
\Input{$Z$ and $\delta$}
\ 

max-heaps $LMax$ and $RMax$\\
min-heaps $LMin$ and $RMin$\\
$bestStat = -\infty$\\
$bestLoc = -1$\\
\For{ $1\le i<\delta$}{
	addToHeaps($LMax$, $LMin$, $Z_i$)\\
}
\For{ $\delta\le i\le n-\delta$}{
	addToHeaps($LMax$, $LMin$, $Z_i$)\\
	mL = getMedian($LMax$, $LMin$)\\
	empty $RMax$ and $RMin$\\
	\For{ $i\le j< i+\delta$}{
		addToHeaps($RMax$, $RMin$, $Z_{j}$)\\
	}
	\For{ $i+\delta\le j\le n$}{
		addToHeaps($RMax$, $RMin$, $Z_j$)\\
		mR = getMedian($RMax$, $RMin$)\\
		\textit{stat} = $\frac{i(j-i)}{j}(mL-mR)^2$\\
		\If{stat > bestStat}{
			\textit{bestStat = stat}\\
			\textit{bestLoc = i}\\
		}
	}
}
\Return{bestLoc}
\label{algo:edmx}
\end{algorithm}

\begin{procedure}[h]
\SetKwInOut{Input}{Parameters}
\SetKwInOut{Output}{Output}
\caption{addToHeaps()}
\Input{$M$, $m$, and $x$}
\ 
\If{$m$ is empty}{
	Add $x$ to $M$\\
}
\If{$m$ isn't empty and $x\le m.top()$}{
	Add $x$ to $M$\\
}
\Else{
	Add $x$ to $m$\\
}
\If{$M.size() > m.size+1$}{
	Move $M.top()$ to $m$\\
}
\If{$m.size() > M.size()+1$}{
	Move $m.top()$ to $M$\\
}
\label{algo:AddToHeap}
\end{procedure}

\begin{procedure}[h]
\SetKwInOut{Input}{Parameters}
\SetKwInOut{Output}{Output}
\caption{getMedian()}
\Input{$M$and $m$}
\Output{The current median}
\If{$m.size() = M.size()+1$}{
	\Return{$m.top()$}
}
\If{$M.size() = m.size()+1$}{
	\Return{$M.top()$}
}
\If{$M.size() = m.size()$}{
	\Return{ $(M.top() + m.top())/2$}
}
\label{algo:GetMedian}
\end{procedure}

%% file: evaluation.tex
\input{smallChange.tex}

\input{tripleEx.tex}

\vspace*{-1mm} 
\section{ Evaluation} \label{evaluation}
\vspace*{-1mm} 

\noindent 
In this section we detail the evaluation methodology and present results 
demonstrating the efficacy, measured in terms of TTD (refer to subsection
\ref{sec:metrics}, of the algorithms presented in the previous section. 
Our experiments show that the presence of anomalies can significantly skew
the TTD of breakout algorithm.

\subsection{Methodology} \label{sec:meth}

The efficacy of {\bf EDM} and {\bf EDM-X} was evaluated using a wide corpus of time 
series data obtained from \underline{production}. The time series corresponded to 
both {\em system} and {\em application} metrics. For example, but not limited to, 
the following metrics were used:

\vspace*{-1mm} 
\begin{dinglist}{114} 
\item System Metrics 
      \vspace*{-1mm} 
      \begin{dinglist}{122}
      \item CPU utilization, Heap usage, Disk writes
            \vspace*{-1mm} 
      \item Time spent in garbage collection
      \end{dinglist}
      \vspace*{-1mm} 
\item Application Metrics
      \vspace*{-1mm} 
      \begin{dinglist}{122} 
      \item Request rate
            \vspace*{-1mm} 
      \item Latency
      \end{dinglist}
\end{dinglist}

\noindent 
In addition to the time series of the metrics mentioned above, we also used 
minutely time series of the stock price of a publicly traded company. Overall, more 
than 20 data sets were used for evaluation. 
Given the velocity, volume, and real-time nature of cloud infrastructure data, 
it is not practical to obtain time series data with ``true" breakouts labeled. 
However, to determine TTD, location of a ``true" breakout is needed. To this
end, for the data sets (obtained from production) we used for evaluation, we 
determined the ``true" breakouts manually and then computed the TTD. 

\subsection{ PELT and E-Divisive}  \label{peltVecp}

\noindent 
Visual analysis serves as the starting point for deriving insights from Big 
Data \cite{Keim10,Wong04,Pitt13}. With the increase in volume in Big Data, 
there has been increasing impetus being given to extreme scale visual 
analytics \cite{Wong12a}. The May 2013 edition of IEEE Computer covered the 
challenges in the realm of Big Data visual analytics \cite{Childs13,Kosara13}. 
However, as mentioned earlier, due to the velocity and volume of cloud data,
visual detection  of breakouts is not practical. Furthermore, sometimes a 
breakout isn't always obvious due to the range of the observed values. This 
is exemplified by \figref{fig:smallChange}. From \figref{fig:smalla} we note 
that there is an anomaly on the left hand side due to which even a 21\% change 
in mean is cannot be detected via visual inspection. However, on zooming in
(in other words, limiting the range of the y-axis), see \figref{fig:smallb}, 
we observe the aforementioned breakout. 

To this end, we first evaluated the {\tt PELT} (Pruned Exact Linear Time) method
by Killick and Haynes \cite{changepointR}. This is a parametric method that can
be used to detect single as well as multiple breakout analysis. In the current
context, we focus only on its properties for estimating a single breakout. This 
method is usually applied by using a log-likelihood function to measure fit, but
as shown in \cite{Killick:2012} the underlying concepts can be extended to a 
number of different measure of fit. One the major benefits of this algorithm is 
its speed, which has been shown to have an expected linear running time.

We also evaluated the {\tt E-Divisive} method \cite{Matteson:2012}. This is a 
non-parametric breakout detection algorithm that is based upon the statistic 
presented in \eqnref{qHat}. Akin to {\tt PELT}, this method can also be used to
estimate multiple breakouts, but we will once again only examine its performance
at identifying a single breakout. However, unlike {\tt PELT}, {\tt E-Divisive} 
is a non-parametric algorithm and makes weak distributional assumptions. Hence, 
{\tt E-Divisive} can be applied in a wider range of settings, such as those 
where one is not certain that {\tt PELT}'s assumptions necessarily hold. On the
other hand, {\tt E-Divisive} has a quadratic running time, which is much slower 
than that of {\tt PELT}.

\input{tablebig.tex}


\begin{figure*}[!t]
\begin{subfigure}{0.5\linewidth}
  \includegraphics[width=1.03\linewidth]{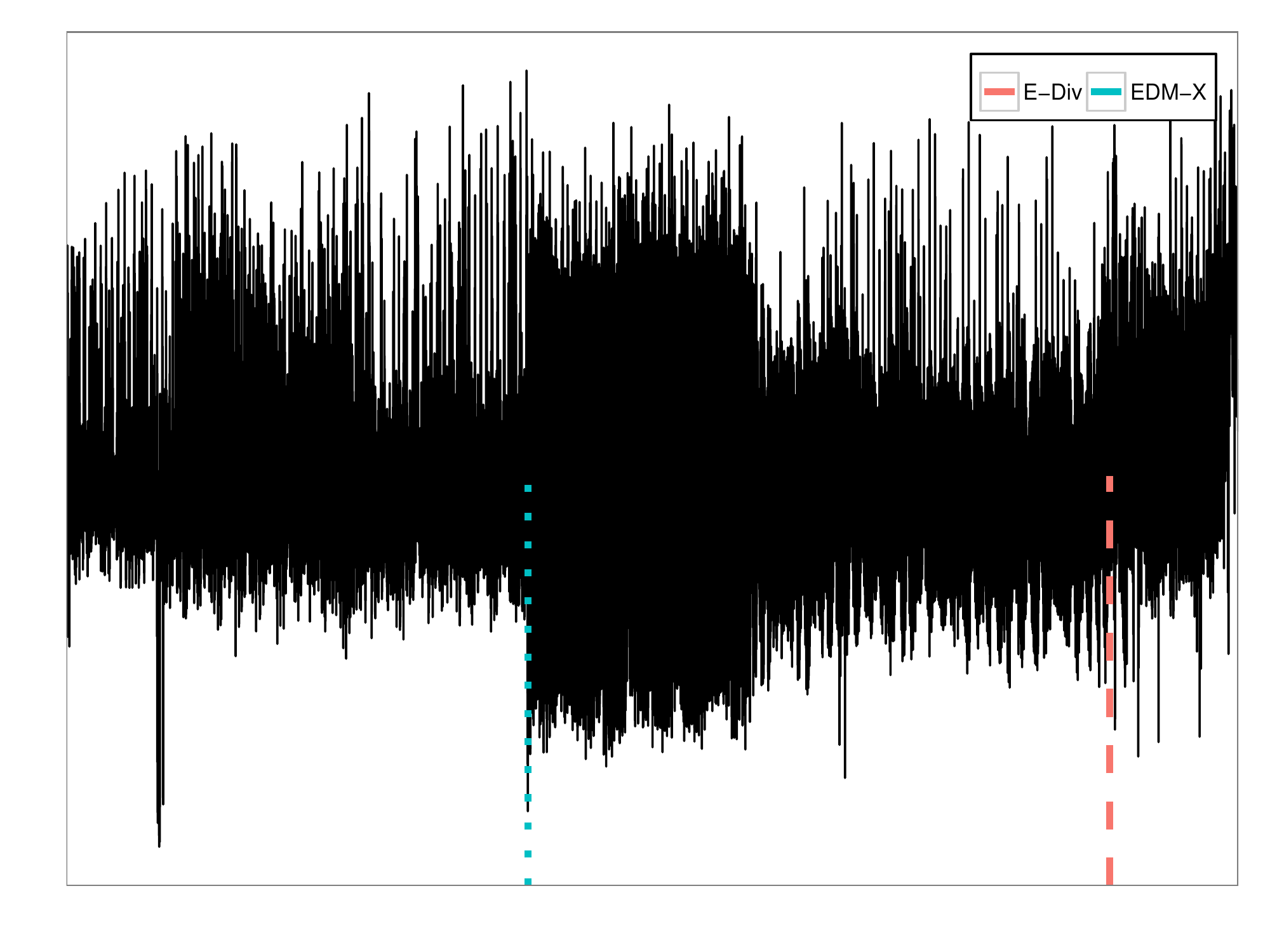}
  \caption{{\tt E-Divisive} and {\bf EDM-X}}
  \label{ex10}
\end{subfigure}
\begin{subfigure}{0.5\linewidth}
  \includegraphics[width=1.03\linewidth]{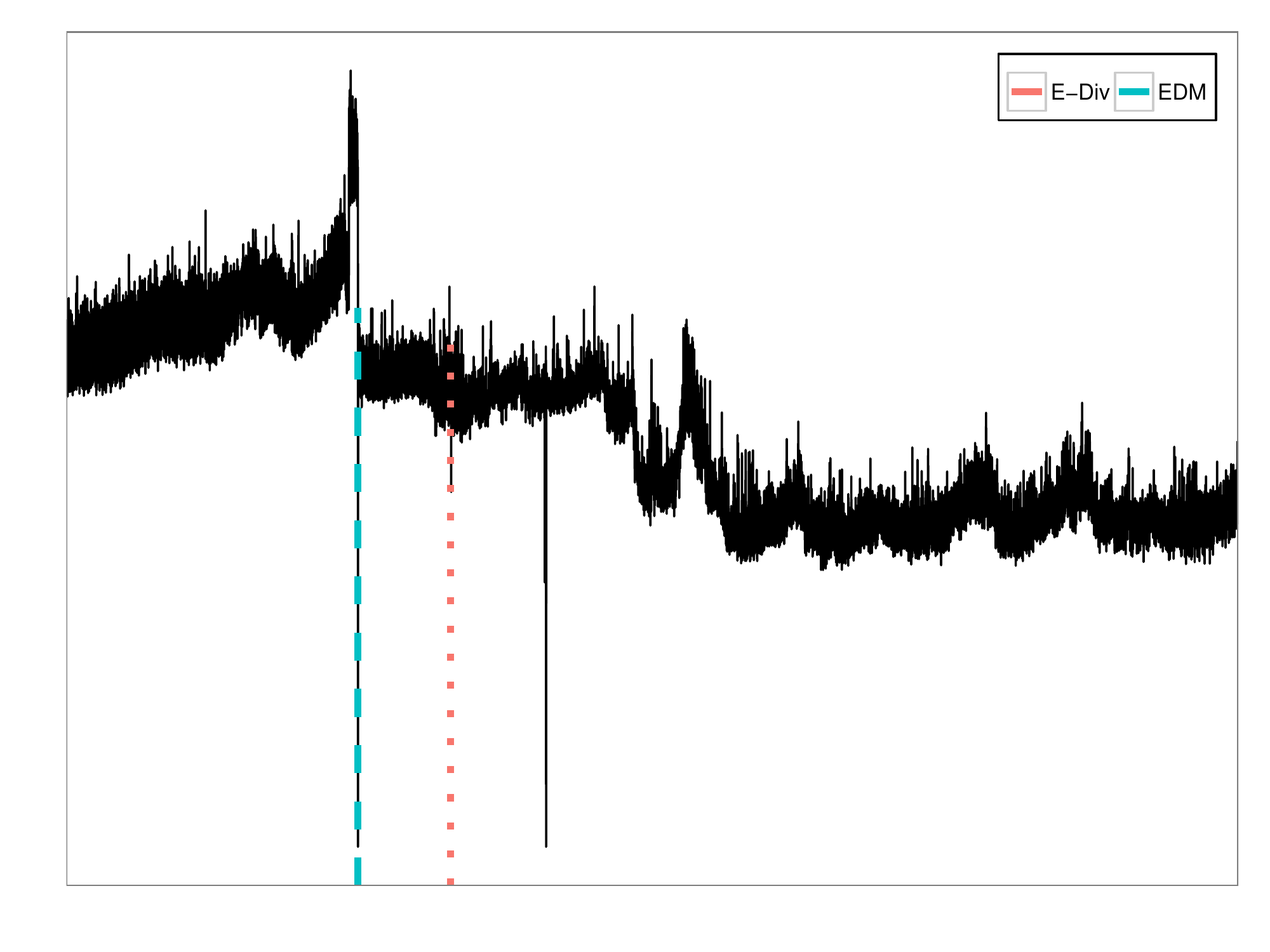}
  \caption{{\tt E-Divisive} and {\bf EDM}}
  \label{ex20}
\end{subfigure}
\vspace*{-3mm}
\caption{Illustration of efficacy of {\bf EDM-X} and {\bf EDM}}
\vspace*{-6mm}
\label{fig:edm_edx}
\end{figure*}

\subsubsection{Data Without Anomalies}

\noindent
First, we applied the {\tt PELT} procedure to the datasets mentioned earlier in
this section. \figref{fig:tripleExa} exemplifies a case wherein the {\tt PELT} 
method is efficient in detecting a breakout. This is further supported by the 
TTD values in column 3 of \tabref{tablebig}. However, since {\tt PELT} makes 
distributional assumptions through its use of likelihood functions, {\tt PELT}'s 
performance suffer -- large TTD value -- when these assumptions do not hold. This
is illustrated by \figref{fig:tripleExb} and column 3 of \tabref{tablebig}.

To address this problem, we used {\tt E-Divisive} to compute breakout location. 
\figref{fig:tripleExb} and column 2 of \tabref{tablebig} show that in almost all 
cases {\tt E-Divisive} results in a smaller TTD. 
Furthermore, since {\tt E-Divisive} is a non-parametric method it can be applied 
to a wider array of settings, especially those where {\tt PELT}'s assumptions are 
not guaranteed to hold. However, although {\tt E-Divisive} is significantly slower
than {\tt PELT} we find this an acceptable trade off because of the decreased TTD
and greater range of applications.

\subsubsection{Data with anomalies}

\noindent 
In the previous section we showed that when a dataset doesn't contain any 
anomalies that both {\tt PELT} and {\tt E-Divisive} can be used to compute
robust estimates locations of a breakout. However, this is not the case in 
the presence of anomalies\footnote{Note that the presence of anomalies in
production cloud is not uncommon \cite{Vallis:2014}.}, as illustrated by
\figref{fig:tripleExc}. 
A common approach to mitigate the effect of anomalies is local smoothing. The 
smoothers we considered were the rolling mean and rolling median. For these 
smoothers, each observation is replaced by either the mean or median of its 
neighboring values. As anomalies can still effect the smoothed values when
calculating the rolling mean, we used the rolling median. 
Although these methods can reduce the impact of anomalies, it can result in 
an increased TTD as seen from columns 4 and 5 of \tabref{tablebig}. Another 
drawback to this approach is that one must choose the size of the neighborhood
to use to calculate the smoothed values. A neighborhood that is too small 
will limit the mitigation of the effect of an anomaly; on the other hand, a
neighborhood one that is too big can potentially smooth the mean changes (a
breakout) in a time series. 
%

Another approach is to remove anomalies before performing breakout analysis.
To this end, we used the S-H-ESD algorithm \cite{Vallis:2014} to automatically 
detect anomalies. Subsequently, the anomalies were removed and breakout was
detected using both {\tt PELT} and {\tt E-Divisive} -- see columns 6 and 7 of 
\tabref{tablebig}. However, we do not consider this an ideal approach as anomaly 
and breakout detection are tightly intertwined. This stems from the fact that 
breakouts can cause normal observations to appear as anomalies, whereas anomalies 
can cause the data to appear to have a different mean. Unlike the local smoothing 
approach preemptive anomaly removal effects both {\tt E-Divisive} and {\tt PELT}. 
Both algorithms become less able to identify a change, as is expected because of 
the relationship between breakout and anomaly detection. 

\input{edmTable.tex}

\subsection{EDM} \label{edmeval}

\noindent 
We next evaluated the efficacy of {\bf EDM}. The TTD values for {\tt E-Divisive},
{\bf EDM-X} and {\bf EDM} are reported in \tabref{edmTable}. Recall that {\bf
EDM} is designed to detect breakouts in an anomaly ``aware" fashion. From the 
table we note that in most cases that TTD values are in the same ballpark as
in the case of {\tt E-Divisive}. In a couple of cases -- Datasets 10 and 20 -- 
both {\bf EDM-X} and {\bf EDM} outperform {\tt E-Divisive} significantly, see 
Figures \ref{ex10} and \ref{ex20}. From \figref{ex10} we note that, unlike {\tt
E-Divisive}, {\bf EDM-X} was able to detect the true location of the change in 
mean. This is due to fact that {\bf EDM-X} was not susceptible to the anomalies 
at the left hand side of the time series. Likewise, from \figref{ex20} we note 
that {\bf EDM} is robust against the anomalies on the right hand side of the 
true location of mean change; hence, {\bf EDM} returned a very accurate estimate 
of the breakout. 

Amongst {\bf EDM-Head} and {\bf EDM-Tail}, the latter seem to perform better 
in most cases. This is desirable from a recency perspective. Only in the case
of Dataset 13 {\bf EDM-Tail} performs significantly worse than {\tt E-Divisive}.



\input{edmTtd.tex}

The {\em Precision, Recall} and {\em F-measure} for both {\bf EDM-X} and {\bf
EDM} is reported in \tabref{edmPR}. From the table we note {\bf EDM-X} has a
higher {\em F-measure} than {\bf EDM-Head} and {\bf EDM-Tail} for the data
sets we used. The approximate p-values obtained using the permutation test
(detailed in \ssecref{sec:permTest}) for each run are tabulated in Table 
\ref{pvalTable}. From the table we see that in some cases the p-value is 
higher than our threshold of 5\%.  




\input{relSpeed.tex}

\input{pvalTable.tex}

Based on our experimental results, we argue for the use of \textbf{EDM} when 
it is suspected that anomalies might be present in a given time series. In
addition, the run time of {\bf EDM-X} and {\bf EDM} is much smaller to that 
of {\tt E-Divisive}, see \figref{fig:relspeed}. 
In our analysis, when performing the permutation test for \textbf{EDM} and 
\textbf{EDM-X}, the maximum number of permutations were always performed. 
However, the implementation of {\tt E-Divisive} in the \textbf{ecp} package 
allows for early termination of the permutation test. Inspite of this, Figure 
\ref{fig:relspeed} shows that \textbf{EDM} and \textbf{EDM-X} are at least 
$2\times$ as fast as {\tt E-Divisive} in almost all cases, and sometimes 
$6\times$ faster.


Even though the \textbf{EDM} and \textbf{EDM-X} algorithms have been shown to be
competitive with {\tt E-Divisive} in the absence of anomalies, and better in the
presence of anomalies, these methods do have their own limitations. For instance,
see \figref{fig:limit}. From the figure we note that \textbf{EDM} reports an
inaccurate breakout estimate. This is attributed to the large number of anomalies
as well as the fact that the anomalies are closely intertwined with the normal
observations.

Another limitation of \textbf{EDM} and \textbf{EDM-X} is that they are both only
able to detect a single breakout. Thus, if more than one breakout exists, it is
unclear which (if any) will be found by {\bf EDM-X} and {\bf EDM}.  
Furthermore, depending on the size and nature of the breakouts, it is possible
for performance to degrade, i.e., TTD may increase. This results from the fact
that both \textbf{EDM} and \textbf{EDM-X} attempt to partition the time series 
into two homogeneous segments.

\input{limit.tex}

%% file: smallChange.tex
\begin{figure*}[!t]
\hspace*{-6mm}
\begin{subfigure}{0.5\linewidth}
  \includegraphics[width=1.03\linewidth]{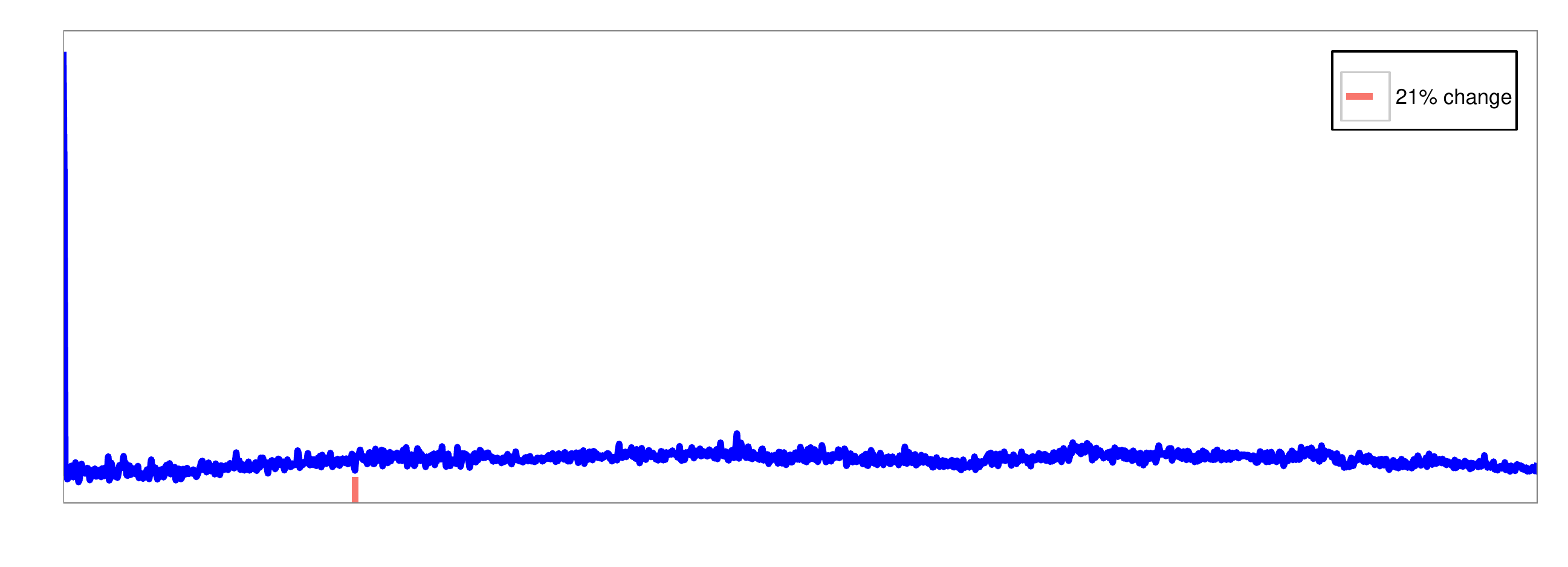}
  \caption{\ }
  \label{fig:smalla}
\end{subfigure}
\begin{subfigure}{0.5\linewidth}
  \includegraphics[width=1.03\linewidth]{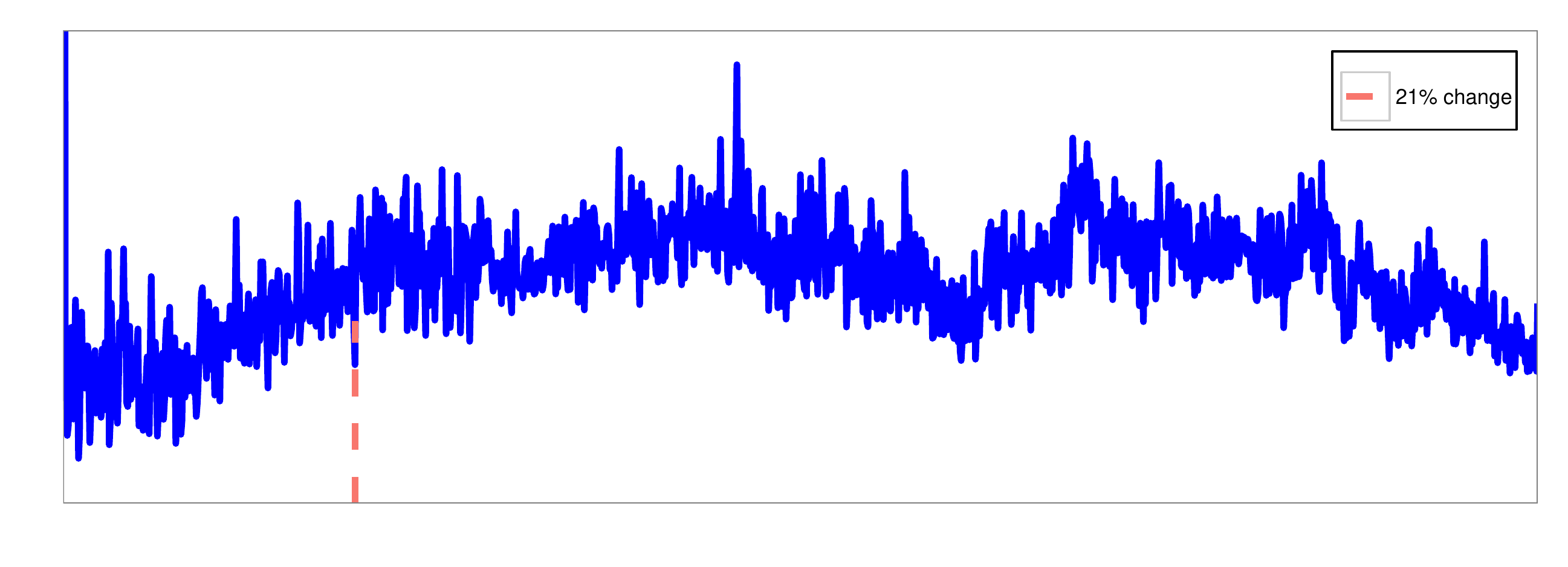}
  \caption{\ }
  \label{fig:smallb}
\end{subfigure}
\vspace{-3mm}
\caption{An example highlighting the limitations of visual detection of breakout(s)}
\vspace{-4mm}
\label{fig:smallChange}
\end{figure*}

%% file: tripleEx.tex
\begin{figure*}[!b]
\vspace{-2mm}
\hspace*{-8mm}
\centering
\begin{subfigure}{0.33\linewidth}
  \includegraphics[width=1.03\linewidth]{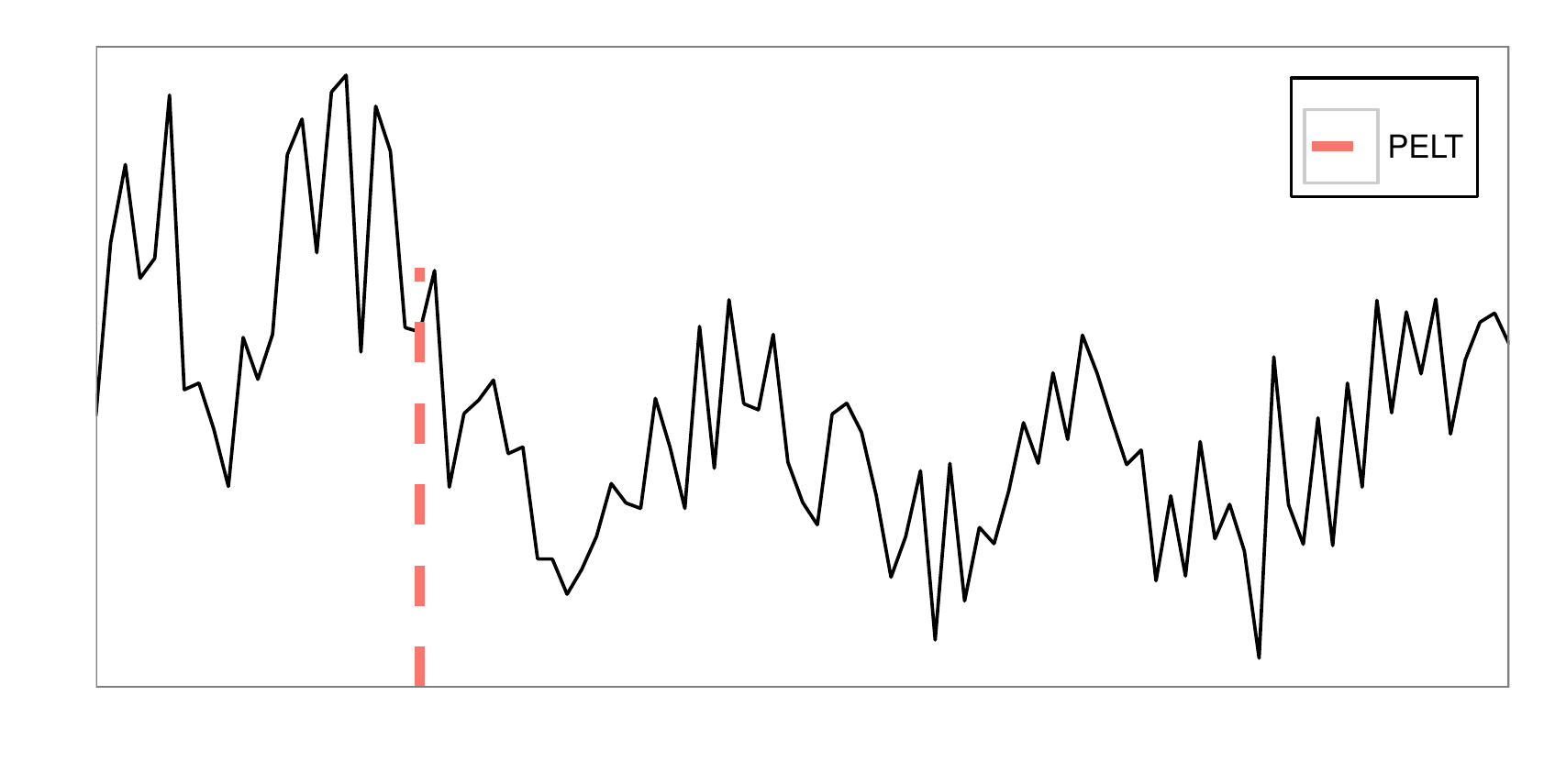}
  \caption{\ }
  \label{fig:tripleExa}
\end{subfigure}
\begin{subfigure}{0.33\linewidth}
  \includegraphics[width=1.03\linewidth]{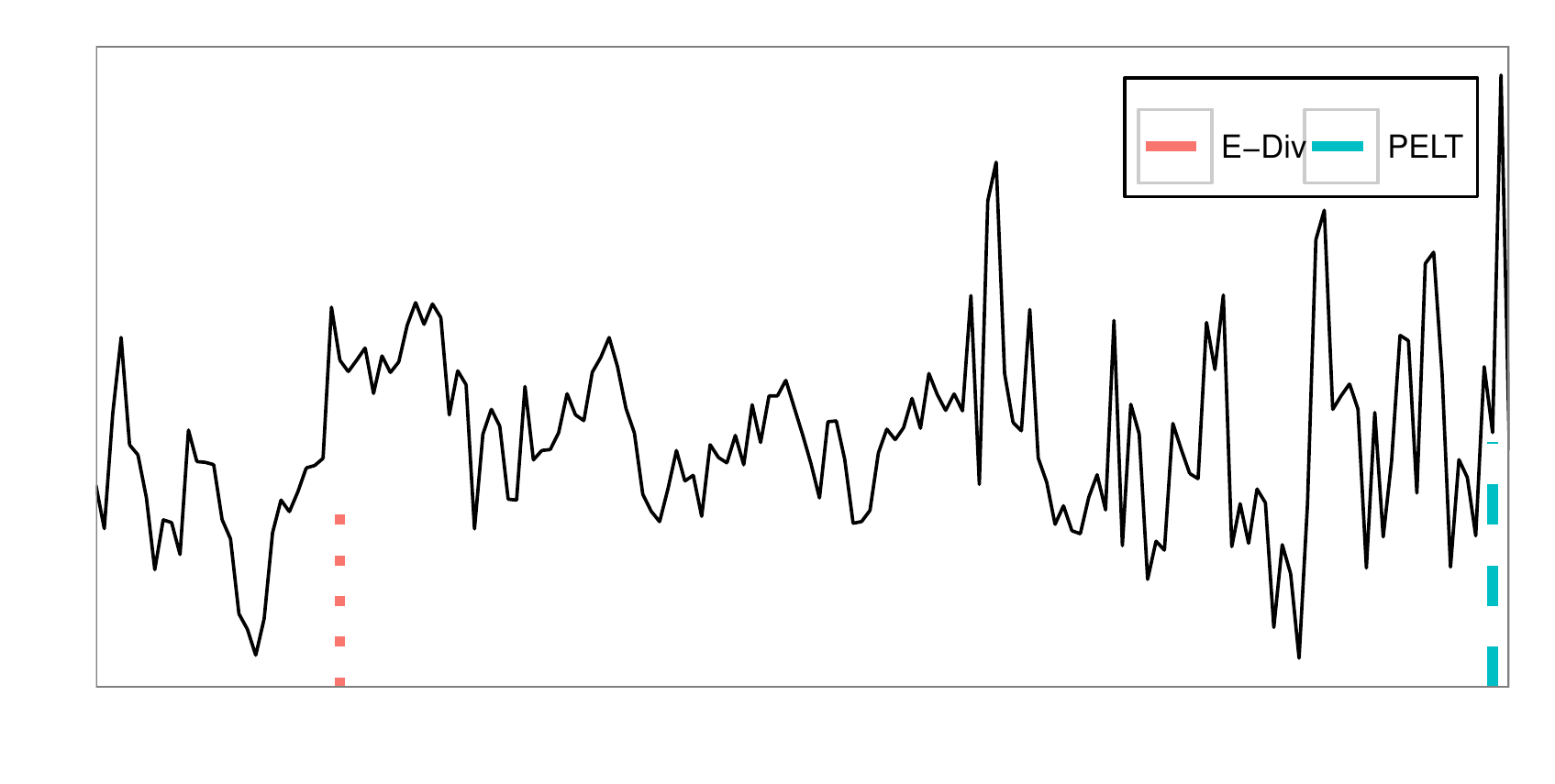}
  \caption{\ }
  \label{fig:tripleExb}
\end{subfigure}
\begin{subfigure}{0.33\linewidth}
  \includegraphics[width=1.03\linewidth]{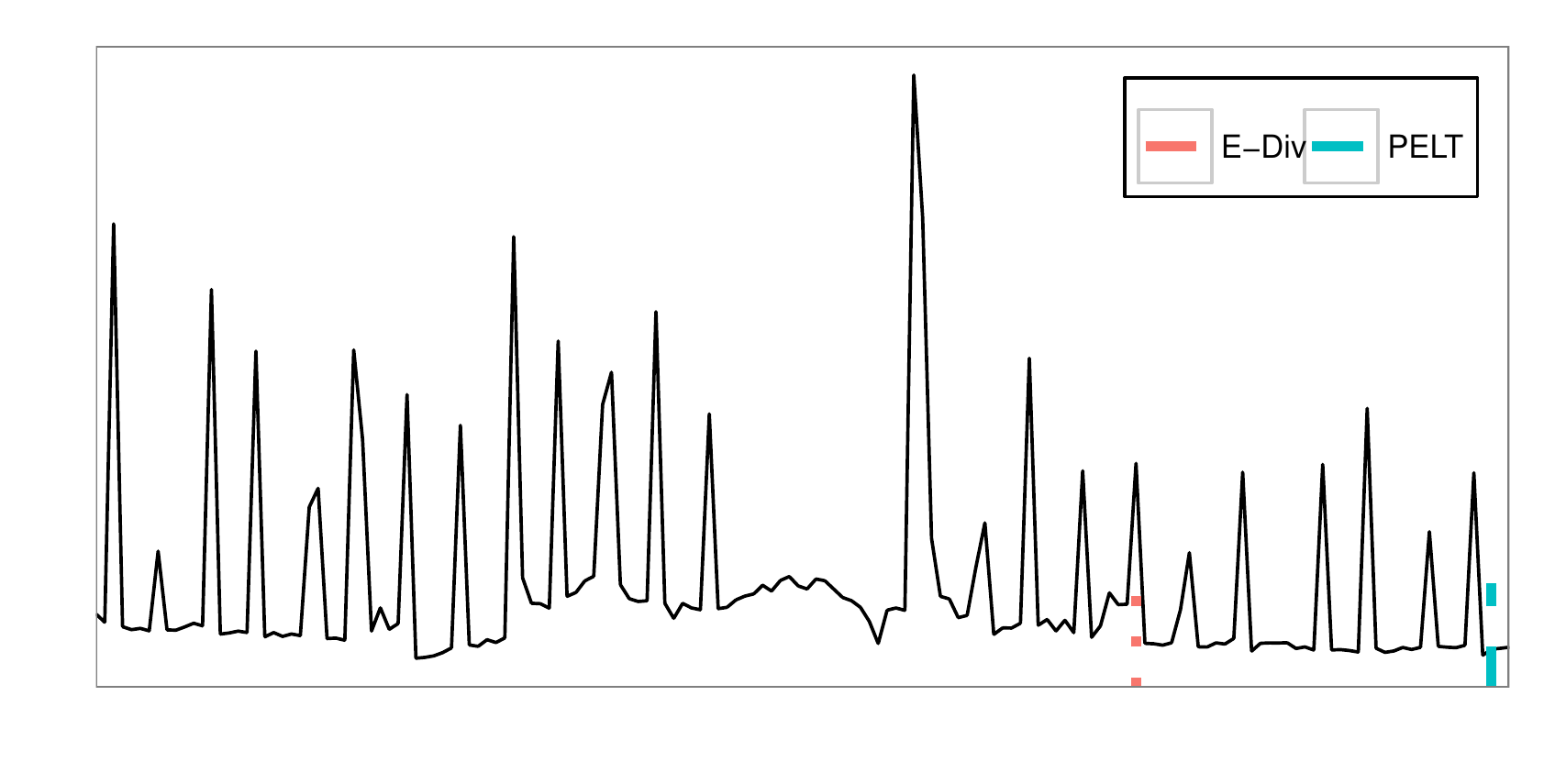}
  \caption{\ }
  \label{fig:tripleExc}
\end{subfigure}
\vspace{-3mm}
\caption{Efficacy of {\tt PELT} and {\tt E-Divisive}}
\label{fig:tripleEx}
\end{figure*}

%% file: tablebig.tex
\begin{table}[!t]
\centering
\resizebox{\linewidth}{!}{
\begin{tabular}{l|rr|rr|rr} \hline 
           & \multicolumn{2}{c|}{\begin{tabular}[c]{@{}c@{}}Raw Data\\ TTD\end{tabular}}    & \multicolumn{2}{c|}{\begin{tabular}[c]{@{}c@{}}Rolling Median\\ TTD\end{tabular}} & \multicolumn{2}{c}{\begin{tabular}[c]{@{}c@{}}Anomalies Removed\\ TTD\end{tabular}} \\ \cline{2-7} 
           & \cellcolor[HTML]{C0C0C0}E-Div & {PELT} & \multicolumn{1}{c}{\cellcolor[HTML]{C0C0C0}E-Div}   & \multicolumn{1}{c|}{PELT}   & \multicolumn{1}{c}{\cellcolor[HTML]{C0C0C0}E-Div}     & \multicolumn{1}{c}{PELT}     \\ \hline \hline
Dataset 1  & \cellcolor[HTML]{C0C0C0}0     		       & 0    			   & \cellcolor[HTML]{C0C0C0}0       			 & 0      		       & \cellcolor[HTML]{C0C0C0}71        		       & 74      		      \\
Dataset 2  & \cellcolor[HTML]{C0C0C0}0     		       & 0    			   & \cellcolor[HTML]{C0C0C0}0       			 & 1      		       & \cellcolor[HTML]{C0C0C0}18        		       & 42      		      \\
Dataset 3  & \cellcolor[HTML]{C0C0C0}2     		       & 1    			   & \cellcolor[HTML]{C0C0C0}1       			 & 1      		       & \cellcolor[HTML]{C0C0C0}1         		       & 1       		      \\
Dataset 4  & \cellcolor[HTML]{C0C0C0}0                         & 6                         & \cellcolor[HTML]{C0C0C0}0                           & 38                          & \cellcolor[HTML]{C0C0C0}0                             & 0                            \\
Dataset 5  & \cellcolor[HTML]{C0C0C0}0                         & 65                        & \cellcolor[HTML]{C0C0C0}1                           & 65                          & \cellcolor[HTML]{C0C0C0}0                             & 65                           \\
Dataset 6  & \cellcolor[HTML]{C0C0C0}2                         & 5                         & \cellcolor[HTML]{C0C0C0}2                           & 5                           & \cellcolor[HTML]{C0C0C0}2                             & 5                            \\
Dataset 7  & \cellcolor[HTML]{C0C0C0}6                         & 7                         & \cellcolor[HTML]{C0C0C0}4                           & 7                           & \cellcolor[HTML]{C0C0C0}1                             & 7                            \\
Dataset 8  & \cellcolor[HTML]{C0C0C0}3                         & 4                         & \cellcolor[HTML]{C0C0C0}2                           & 4                           & \cellcolor[HTML]{C0C0C0}3                             & 4                            \\
Dataset 9  & \cellcolor[HTML]{C0C0C0}9                         & 8                         & \cellcolor[HTML]{C0C0C0}6                           & 8                           & \cellcolor[HTML]{C0C0C0}8                             & 15                           \\
Dataset 10 & \cellcolor[HTML]{C0C0C0}14113                     & 15                        & \cellcolor[HTML]{C0C0C0}14114                       & 16                          & \cellcolor[HTML]{C0C0C0}14113                         & 15                           \\
Dataset 11 & \cellcolor[HTML]{C0C0C0}0                         & 1                         & \cellcolor[HTML]{C0C0C0}0                           & 4                           & \cellcolor[HTML]{C0C0C0}-                             & -                            \\
Dataset 12 & \cellcolor[HTML]{C0C0C0}0                         & 1                         & \cellcolor[HTML]{C0C0C0}1                           & 1                           & \cellcolor[HTML]{C0C0C0}1                             & 1                            \\
Dataset 13 & \cellcolor[HTML]{C0C0C0}45                        & 2                         & \cellcolor[HTML]{C0C0C0}45                          & 2                           & \cellcolor[HTML]{C0C0C0}-                             & -                            \\
Dataset 14 & \cellcolor[HTML]{C0C0C0}0                         & 1                         & \cellcolor[HTML]{C0C0C0}1                           & 2                           & \cellcolor[HTML]{C0C0C0}0                             & 1                            \\
Dataset 15 & \cellcolor[HTML]{C0C0C0}1                         & 1590                      & \cellcolor[HTML]{C0C0C0}0                           & 1590                        & \cellcolor[HTML]{C0C0C0}-                             & -                            \\
Dataset 16 & \cellcolor[HTML]{C0C0C0}0                         & 1                         & \cellcolor[HTML]{C0C0C0}0                           & 1                           & \cellcolor[HTML]{C0C0C0}-                             & -                            \\
Dataset 17 & \cellcolor[HTML]{C0C0C0}2                         & 263                       & \cellcolor[HTML]{C0C0C0}1                           & 263                         & \cellcolor[HTML]{C0C0C0}1681                          & 1733                         \\
Dataset 18 & \cellcolor[HTML]{C0C0C0}1                         & 0                         & \cellcolor[HTML]{C0C0C0}2                           & 0                           & \cellcolor[HTML]{C0C0C0}2                             & 0                            \\
Dataset 19 & \cellcolor[HTML]{C0C0C0}2                         & 61                        & \cellcolor[HTML]{C0C0C0}1                           & 61                          & \cellcolor[HTML]{C0C0C0}105                           & 108                          \\
Dataset 20 & \cellcolor[HTML]{C0C0C0}4479                      & 5607                      & \cellcolor[HTML]{C0C0C0}4476                        & 5607                        & \cellcolor[HTML]{C0C0C0}4479                          & 5607                         \\
Dataset 21 & \cellcolor[HTML]{C0C0C0}27                        & 349 & \cellcolor[HTML]{C0C0C0}41                          & 13 & \cellcolor[HTML]{C0C0C0}-                             & - \\ 
Dataset 22 & \cellcolor[HTML]{C0C0C0}0                      & 0                      & \cellcolor[HTML]{C0C0C0}3                        & 19                        & \cellcolor[HTML]{C0C0C0}4                          & 4                         \\
Dataset 23 & \cellcolor[HTML]{C0C0C0}4                      & 1                      & \cellcolor[HTML]{C0C0C0}15                        & 15                        & \cellcolor[HTML]{C0C0C0}-                          & -                         \\
Dataset 24 & \cellcolor[HTML]{C0C0C0}32                      & 44                      & \cellcolor[HTML]{C0C0C0}17                        & 0                        & \cellcolor[HTML]{C0C0C0}1                          & 1                         \\
Dataset 25 & \cellcolor[HTML]{C0C0C0}0                      & 6                      & \cellcolor[HTML]{C0C0C0}18                        & 89                        & \cellcolor[HTML]{C0C0C0}5                          & 4                         \\ \hline\hline
\end{tabular}}
\caption{TTD for the {\tt E-Divisive} and {\tt PELT} methods when applied to raw and rolling median time series}
\vspace*{-10mm}
\label{tablebig}
\end{table}

%% file: edmTable.tex
\begin{table}[h]
\centering
\resizebox{\linewidth}{!}{
\begin{tabular}{l|r|r|r|r} \hline
           & E-Div                         & {\bf EDM-X}            & {\bf EDM-Head} & {\bf EDM-Tail} \\ \hline\hline
Dataset 1  & \cellcolor[HTML]{C0C0C0}0     & 0                      & 4    & 6    \\
Dataset 2  & \cellcolor[HTML]{C0C0C0}0     & {64} & 84   & 0    \\
Dataset 3  & \cellcolor[HTML]{C0C0C0}2     & {12} & 2    & 0    \\
Dataset 4  & \cellcolor[HTML]{C0C0C0}0     & 0                      & 3    & 2    \\
Dataset 5  & \cellcolor[HTML]{C0C0C0}0     & 0                      & 5    & 1    \\
Dataset 6  & \cellcolor[HTML]{C0C0C0}2     & 1                      & 0    & 0    \\
Dataset 7  & \cellcolor[HTML]{C0C0C0}6     & 6                      & 26   & 6    \\
Dataset 8  & \cellcolor[HTML]{C0C0C0}3     & 1                      & 69   & 11   \\
Dataset 9  & \cellcolor[HTML]{C0C0C0}9     & 3                      & 8    & 0    \\
Dataset 10 & \cellcolor[HTML]{C0C0C0}14113 & 8                      & 1489 & 43   \\
Dataset 11 & \cellcolor[HTML]{C0C0C0}0     & 66                     & 5    & 1    \\
Dataset 12 & \cellcolor[HTML]{C0C0C0}0     & 0                      & 47   & 2    \\
Dataset 13 & \cellcolor[HTML]{C0C0C0}45    & 215                    & 246  & 1332 \\
Dataset 14 & \cellcolor[HTML]{C0C0C0}0     & 78                     & 5    & 0    \\
Dataset 15 & \cellcolor[HTML]{C0C0C0}1     & 3                      & 9    & 4    \\
Dataset 16 & \cellcolor[HTML]{C0C0C0}0     & 268                    & 89   & 95   \\
Dataset 17 & \cellcolor[HTML]{C0C0C0}2     & 1                      & 122  & 1    \\
Dataset 18 & \cellcolor[HTML]{C0C0C0}1     & 26                     & 0    & 0    \\
Dataset 19 & \cellcolor[HTML]{C0C0C0}2     & 27                     & 4    & 1    \\
Dataset 20 & \cellcolor[HTML]{C0C0C0}4479  & 183                    & 55   & 3    \\
Dataset 21 & \cellcolor[HTML]{C0C0C0}27    & 70                     & 204  & 78   \\
Dataset 22 & \cellcolor[HTML]{C0C0C0}0  & 0                    & 34   & 4    \\
Dataset 23 & \cellcolor[HTML]{C0C0C0}4  & 19                    & 0   & 4    \\
Dataset 24 & \cellcolor[HTML]{C0C0C0}32  & 143                    & 47   & 3    \\
Dataset 25 & \cellcolor[HTML]{C0C0C0}0  & 11                    & 6   & 2    \\ \hline\hline
\end{tabular}}
\caption{TTD for various nonparametric breakout procedures. {\bf EDM-Head} 
         and {\bf EDM-Tail} refer to the {\bf EDM} algorithm when the $\delta$ 
         between distance observations is chosen according to the Figures 
         \ref{headTail}(a) and \ref{headTail}(b) respectively}
\vspace*{-6mm}
\label{edmTable}
\end{table}

%% file: edmTtd.tex
\begin{table}[!b]
\vspace*{-3mm}
\centering
\begin{tabular}{l|r|r|r} \hline
           & {\bf EDM-X}            & {\bf EDM-Head} & {\bf EDM-Tail} \\ \hline\hline
Precision  & 0.8400             &  0.9048    & 0.9048    \\
Recall     & 1                      &  0.8261  & 0.8261 \\
F-Measure  & 0.9130 & 0.8636& 0.8636\\ \hline\hline
\end{tabular}
\caption{Precision, recall, and F-Measure for {\bf EDM-X} and \textbf{EDM}}
\vspace*{-6mm}
\label{edmPR}
\end{table}


%% file: relSpeed.tex
\begin{figure}[!b]
\hspace*{-4mm}
\includegraphics[width=1.05\linewidth]{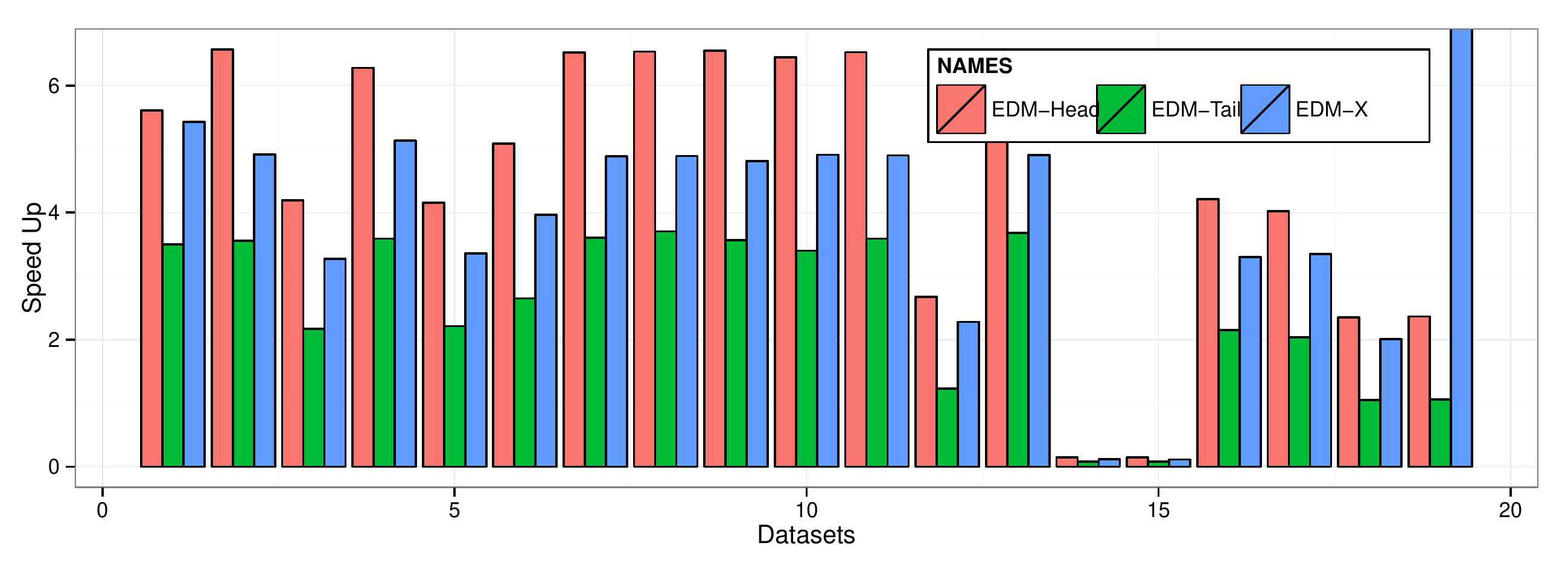}
\vspace{-6mm}
\caption{Speedup of \textbf{EDM} and \textbf{EDM-X} relative to {\tt E-Divisive}}
\label{fig:relspeed}
\end{figure}

%% file: pvalTable.tex
\begin{table}[h]
\centering
\begin{tabular}{l|r|r|r} \hline
           & {\bf EDM-X}            & {\bf EDM-Head} & {\bf EDM-Tail} \\ \hline\hline
Dataset 1  &  0.005                  & 0.130   & 0.115    \\
Dataset 2  &  0.005                  & 0.005   & 0.005    \\
Dataset 3  &  0.005                  & 0.005   & 0.005    \\
Dataset 4  &  0.005                  & 0.005   & 0.005    \\
Dataset 5  &  0.005                  & 0.100   & 0.050    \\
Dataset 6  &  0.005                  & 0.005   & 0.005    \\
Dataset 7  &  0.005                  & 0.005   & 0.005    \\
Dataset 8  &  0.005                  & 0.035   & 0.120   \\
Dataset 9  &  0.005                  & 0.005   & 0.015    \\
Dataset 10 &  0.005                  & 0.005   & 0.005   \\
Dataset 11 &  0.005                  & 0.005   & 0.005    \\
Dataset 12 &  0.005                  & 0.015   & 0.010    \\
Dataset 13 &  0.005                  & 0.010   & 0.010 \\
Dataset 14 &  0.005                  & 0.005   & 0.005    \\
Dataset 15 &  0.005                  & 0.005   & 0.005    \\
Dataset 16 &  0.005                  & 0.005   & 0.005   \\
Dataset 17 &  0.005                  & 0.005   & 0.005    \\
Dataset 18 &  0.005                  & 0.005   & 0.005    \\
Dataset 19 &  0.005                  & 0.085   & 0.020    \\
Dataset 20 &  0.005                  & 0.005   & 0.005    \\
Dataset 21 &  0.005                  & 0.925   & 0.990   \\
Dataset 22 &  0.005                  & 0.020   & 0.985    \\
Dataset 23 &  0.005                  & 0.005   & 0.005    \\
Dataset 24 &  0.005                  & 0.025   & 0.030    \\
Dataset 25 &  0.005                  & 0.005   & 0.005    \\ \hline\hline
\end{tabular}
\caption{Approximate p-values obtained from permutation test (detailed in \ssecref{sec:permTest})}
\vspace*{-6mm}
\label{pvalTable}
\end{table}

%% file: limit.tex
\begin{figure}[!h]
\centering
\includegraphics[width=1.05\linewidth]{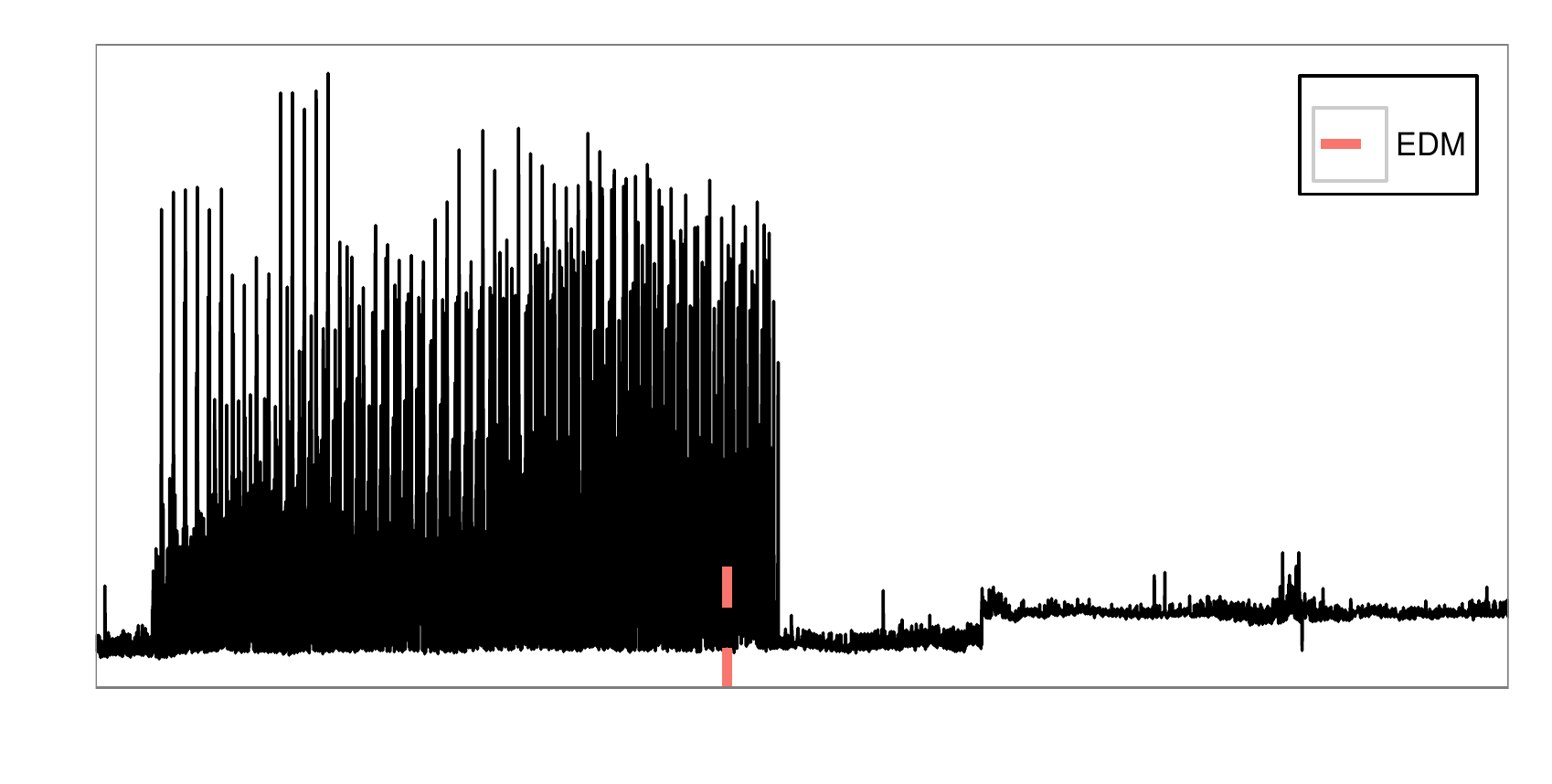}
\vspace{-8mm}
\caption{An example illustrating limitations of \textbf{EDM}}
\vspace{-5mm}
\label{fig:limit}
\end{figure}

%% file: relWork.tex
\section{ Related Work} \label{sec:rel}

Breakout detection has been research in a wide variety of fields owing to the 
different applications. In this section we present a brief overview of prior
work in breakout detection in statistics, finance, medicine and signal 
processing.

\noindent 
As mentioned earlier, {\em breakout} is referred to as a {\em changepoint} in 
statistics. Changepoint detection has been researched in statistics for over 
five decades \cite{Page:1955,Brodsky:1993,Csorgo:1997,Chen:2011}. These come 
in two flavors: parametric and non-parametric. Many of the existing parametric
methods assume that the underlying distribution belongs to the exponential family
\cite{Chen:2011}. There has been recent research in detecting changes with heavy 
tailed distributions \cite{Raimondo:2004}. Many of these approaches make use of 
limiting distributions obtained from Extreme Value Theory \cite{Gumbel:2012}. In
cases where it is difficult or impossible to prove that the data adheres to
parametric assumptions non-parametric approaches provide an alternative solution.
These methods place less restrictive assumptions on the data and can thus be used
more widely in general; however, due to the weaker assumptions, these methods 
are less powerful than their parametric counterparts \cite{Zou:2014}. Although 
most of the prior researched centered around detecting changes in mean, detecting 
changes in variance (with known/unknown mean value) has garnered some attention 
\cite{Chen:1997,Adegboye:1989,Whitcher:2002}. 


Tsay \cite{Tsay:1988} presents an approach to detect changes in mean of an ARMA 
model in the presence of anomalies. Unlike {\bf EDM}, the approach employs a 
two staged process that first removes the anomalies and then carry out breakout 
analysis. Another approach to handle anomalies during breakout detection is to 
assume that the data follows a heavy tailed distribution \cite{Loretan:1994} and 
thus large values become less uncommon \cite{Embrechts:1999}.

\subsubsection{Parametric Analysis}

\noindent 
The parametric algorithms used to perform breakout analysis assume that the
observed distributions belong to a family of distributions $\mathcal F_\Theta =
\{F_\theta: \theta\in\Theta\}$, such that each member of the family can be
uniquely identified by its parameter value. Once the class of distributions 
has been specified, parametric methods attempt to detect changes in the 
parameter value. Specifically, these approaches usually attempt to maximize 
a likelihood. For example, Carlin et al.,\ \cite{Carlin:1992}, Lavielle and
Teyssi\`ere \cite{Lavielle:2006} employ this approach. These papers however,
assume a Gaussian distribution. An extension of this to select methods of
the exponential family \cite{Erling:1970} is supported in the {\tt changepoint} 
R package \cite{changepointR}. 

\subsubsection{Non-parametric Analysis}


\noindent 
A very common approach is to perform density estimation \cite{Kawahara:2012}. 
Although density estimation seems like a natural approach, other ideas have 
been shown to yield satisfactory results. For example, Lung-Yut-Fong et al.\, 
\cite{Lung:2011} perform analysis by working with rank statistics; Matteson 
and James \cite{Matteson:2012} present an approach based upon Euclidean
distances.

\subsection{Finance}

\noindent 
One of the more popular application areas of breakout detection is finance 
\cite{Kirkpatrick:2010,Grimes:2012,Edwards:2012}. In this regard, models are
regularly used to analyze return and stock price data. It is often assumed 
that the model parameters remain constant over the observed period. However, 
if the parameters are mostly time varying, the obtained results are likely 
to become out-of-date and consequently may not be robust \cite{Grimes:2012}. 
Explicit examples of trading strategies that make use of breakout detection 
can be found in \cite{Grimes:2012} which rely on historical analysis, charts 
and familiarity with the market. 

The ARCH model of Engle \cite{Engle:1982} and its various generalizations are 
very often used to model the returns for a number of financial instruments. 
Franses and Ghijsels \cite{Franses:1999} present a method for fitting GARCH 
models to financial data that may have additive outliers. In a similar vein,
in \cite{Matteson:2013}, Matteson and James presented an approach that only 
requires a few mild statistical conditions to hold and doesn't rely on any 
back testing. Regardless of the strategy, both works show that acknowledging 
the existence of breakouts can increase profits, or better yet, change would 
be losses into gains.

\subsection{Medical Applications}

\noindent 
Breakout detection also has applications in medicine. For example, Grigg et
al.\ describe the use of the cumulative sum (CUSUM) chart, RSPRT (resetting
sequential probability ratio test), and FIR (fast initial response) CUSUM
to detect improvements in a process as well as detecting deterioration in
a medical setting. 
In genetics, array comparative genomic hybridization is used to record DNA copy
numbers.  Changes in the DNA copy number can indicate a portion of a gene that 
may be effected by cancer or some other abnormal feature. Thus, detecting 
breakout in this setting \cite{Olshen:2004,Bleakley:2011} can provide insights 
about future medical research. 
Breakout analysis also finds application in segmentation of electroencephalogram
(EEG). An EEG is a measure of the brain's electrical activity which is recoded by 
electrodes on the subject's scalp. EEGs can be used in the process of diagnosing 
disorders such as epilepsy and insomnia, since such disorders cause clear changes
in the EEG readings. Breakout procedures have been suggested as a way to remove 
the human bias in the analysis of such data \cite{Barlow:1981,Kaplan:2000}. Other
application areas include studying breast cancer survival rates \cite{Contal:1999}, 
analysis of fMRI data \cite{Robinson:2010}, and many more \cite{Chen:2011}.

\subsection{Signal Processing}

\noindent 
Breakouts detection has been researched in the field of signal processing (and 
others such as, but not limited to, computer vision, image processing) but is 
usually referred to {\em edge detection} or {\em jump detection} \cite{Willsky76,
Segen80,Basseville81,benveniste-pi564}. In \cite{Basseville:1988}, Basseville 
presented a survey of techniques to detect changes in signals and systems; Ziou
and Tabbone present an overview of edge detection techniques in \cite{Ziou:1998}. 
In the context of dynamic systems, Tugnait presented techniques to detect changes 
in \cite{Tugnait82}.

In \cite{Jackson:2005}, Jackson et al.\ presented an algorithm for optimal
partitioning of data on an interval. The algorithm was subsequently enhanced
by Killick et al.\ \cite{Killick:2012} to detect breakouts with an expected 
linear running time. 

%
%

%% file: conclusion.tex
\section{Conclusions} \label{sec:conclusion}

\noindent
In this paper, we proposed a novel statistical technique, called {\bf  E-Divisive 
with Medians (EDM)}, to automatically detect breakouts in cloud data. Unlike 
the existing techniques for breakout detection, {\bf EDM} is robust against 
the presence of anomalies. {\bf EDM} employs E-statistics \cite{Szekely:2013} 
to detect divergence in mean. Note that, in general, {\bf EDM} can also be used
detect change in distribution in a given time series. Further, {\bf EDM} uses
robust statistical metrics, viz., median, and estimates the statistical
significance of a breakout through a permutation test. 
We used \underline{production} data and to evaluate the efficacy of {\bf EDM}
and reported Precision, Recall and F-measure to demonstrate the same. {\bf EDM}
is $3.5\times$ faster than the state-of-the-art technique for breakout detection
and is being currently used on a daily basis at Twitter.

As future work, we intend to extend {\bf EDM} to support breakout detection in
the presence of seasonality. Further, we plan to explore data transformation
techniques to address the limitations mentioned in \secref{evaluation}.

%% file: appendix.tex
\begin{figure*}[!t] 
\begin{multicols}{4}
\includegraphics[scale=.35]{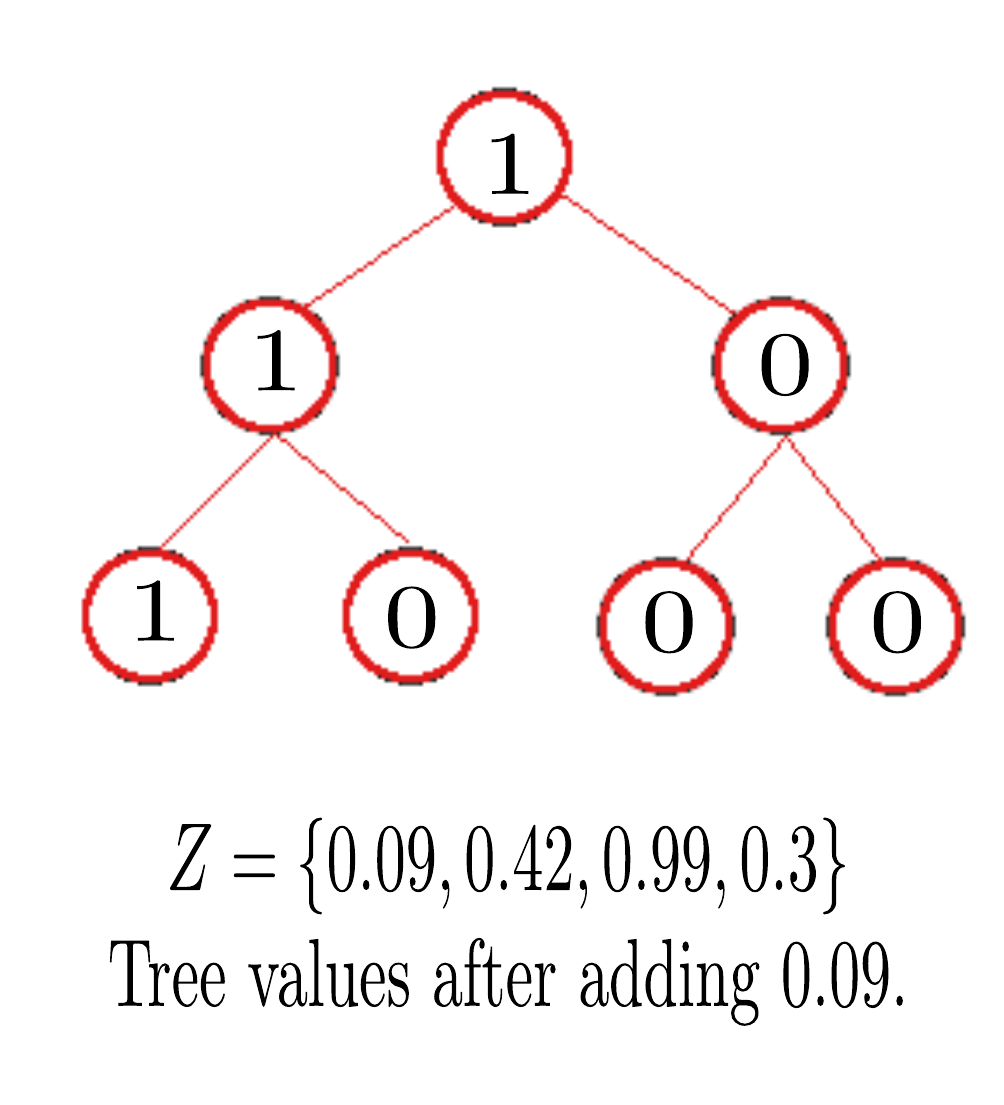} 
\columnbreak
\hspace*{-7mm}
\includegraphics[scale=.35]{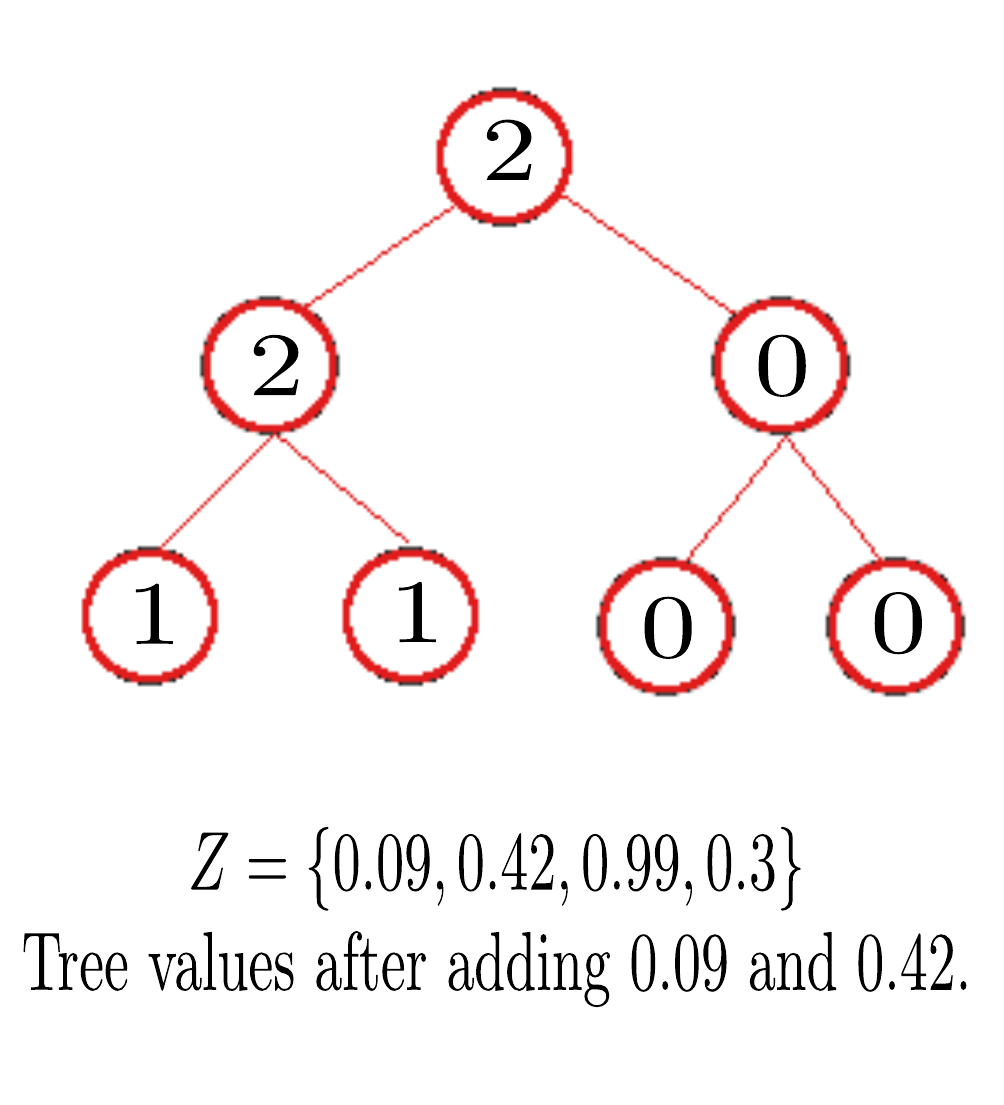} 
\columnbreak 
\hspace*{-4mm}
\includegraphics[scale=.35]{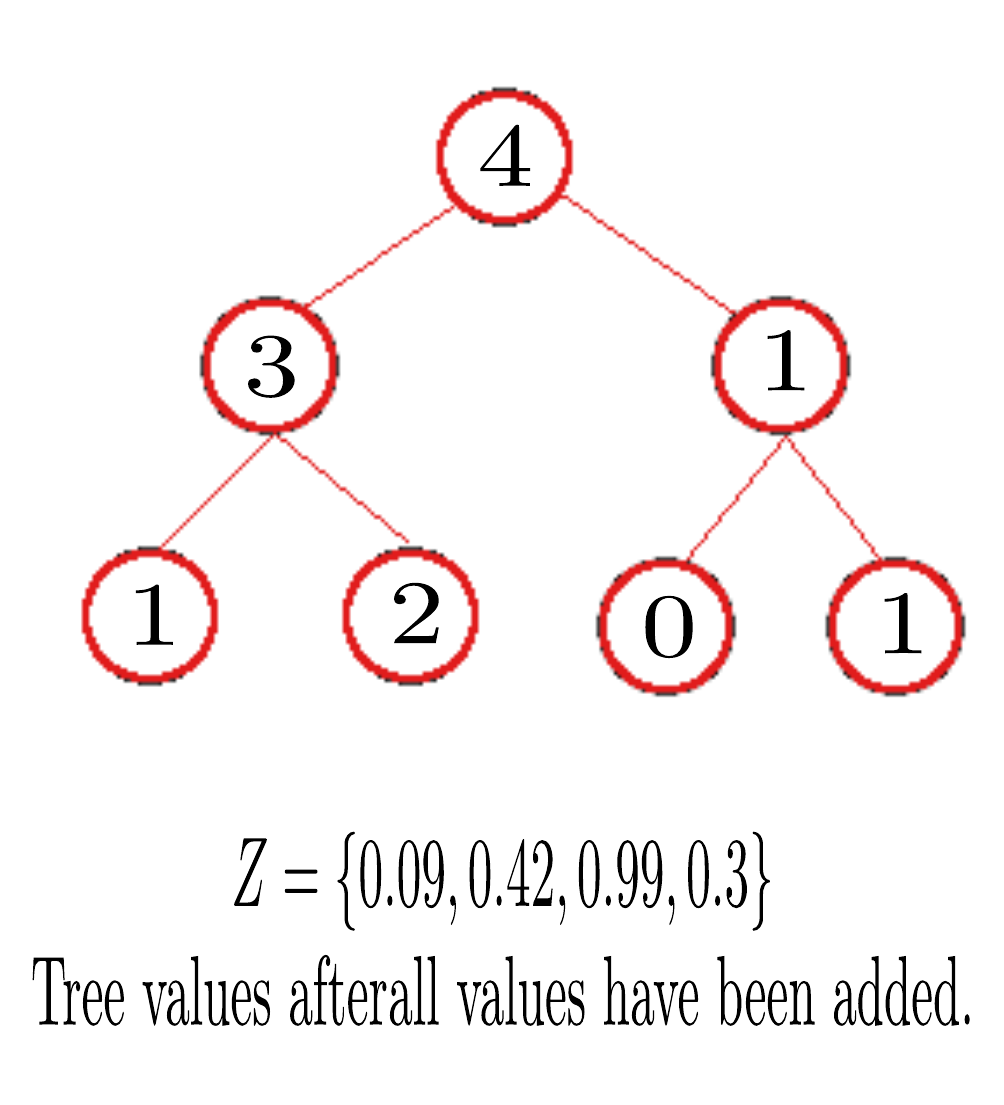}
\columnbreak 
\hspace*{-4mm}
\includegraphics[scale=.35]{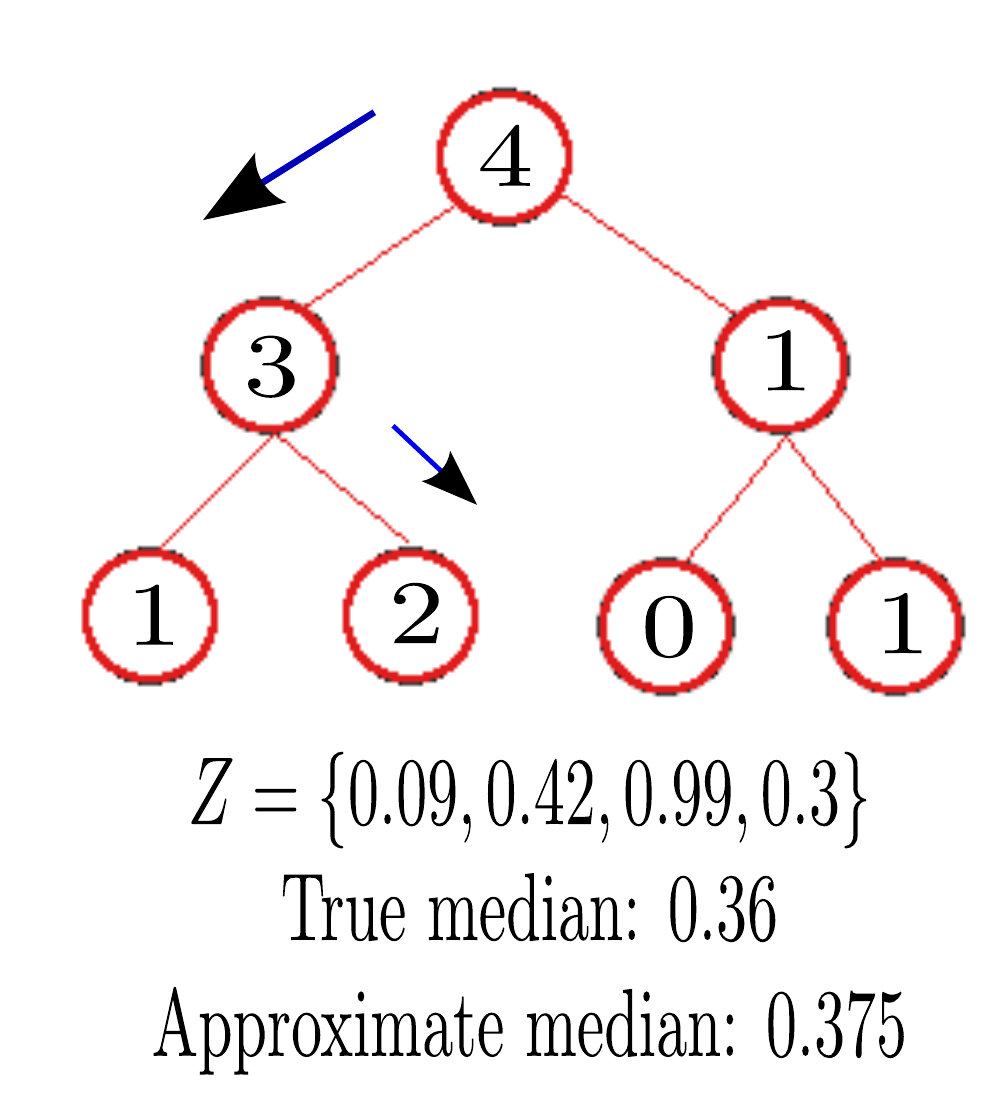} 
\end{multicols}
\vspace*{-7mm}
\caption{Illustration of the use of interval trees to determine approximate median} 
\vspace*{-3mm}
\label{fig:intervalTree} 
\end{figure*}

\section{Appendix} \label{sec:appendix}
\vspace*{-1mm}

\noindent 
In this appendix we present an in-depth description of the details necessary
to implement both \textbf{EDM} and \textbf{EDM-X}, as well as the interval 
tree used to calculate approximate medians. All of these algorithms assume 
that our time series values lie within the interval $[0,1]$. 
Thus if $M = \max\{Z_i: 1\le i\le n\}$ an $m=\min\{Z_i: 1\le i \le n\}$ we 
transform our observations according to the following linear function

\vspace*{-3mm}
\begin{equation*}
f(x) = \frac{x-m}{M-m}.
\end{equation*}
\vspace*{-1mm}

\noindent 
It should be noted that this transformation only scales the value of our 
approximate (or true) median by a value of $\frac{1}{M-m}$.

\subsection{Interval Trees}\label{sec:intervalTree}

\noindent 
In this subsection we detail how interval trees are used by {\bf EDM} and 
{\bf EDM-X}. 
%
%
Our interval tree is a complete binary tree with $2^D$ leaf nodes, where $D$ 
is the user specified depth. The $i$th leaf node represents the interval
$\left[\frac{i-1}{2^D},\frac{i}{2^D}\right)$, 
except for the $2^D$th interval which is a closed interval, instead of a half-open 
interval. Each internal node corresponds to the union of the intervals of its 
children. Thus, the root represents the interval $[0,1]$. In this data structure
each node will contain a count of the number of observations that lie within 
its interval.

Owing to the nature of the tree, one can find an approximate median in $\mathcal
O(D)$ time. One can find a value $m$, such that $K=\lceil\frac{n}{2}\rceil$ of
our observations are less that or equal to $m$ in the following manner:
Starting at the root node compare the value of its left child with $K$. If its
value is larger than $K$, move to that node. On the other hand, if $K$ is larger, 
subtract the value of the left node from $K$ and move to the right child. This 
procedure is continued until a leaf node is reached; then, the midpoint of the
leaf's corresponding interval is returned. However, if at some point an internal
node is reached who's  value is equal to $K$, the following is carried out:
Let $a$ and $b$ be the values of the left and right children respectively, and 
$x,y$ the midpoints of their corresponding intervals. The following is returned:

$$ \frac{1}{a+b}\left(a\times x + b\times y\right) $$

\noindent
The major benefit of using an interval tree to obtain an approximate median
instead of finding the true median is that the data structure can be updated
efficiently and does not require sorting. Furthermore, from our experiments 
we have found that the relative difference between the true median and the 
approximation to be below 10\%. \figref{fig:intervalTree} illustrates how to 
update the tree as well as how to an approximate median.